\documentclass[11pt]{article}%
\usepackage{amsmath}
\usepackage{amsthm}
\usepackage{amsfonts}
\usepackage{amssymb}
\usepackage{graphicx}
\usepackage{fullpage}%
\providecommand{\U}[1]{\protect\rule{.1in}{.1in}}
\newtheorem{theorem}{Theorem}

\newtheorem{claim}[theorem]{Claim}

\newtheorem{conjecture}[theorem]{Conjecture}
\newtheorem{corollary}[theorem]{Corollary}

\newtheorem{definition}[theorem]{Definition}

\newtheorem{lemma}[theorem]{Lemma}

\newtheorem{proposition}[theorem]{Proposition}

\newtheorem*{numlessthm}{Theorem}
\begin{document}

\title{The Power of Unentanglement}
\author{Scott Aaronson\thanks{To whom correspondence should be addressed. \ Email:
aaronson@csail.mit.edu.}\\MIT
\and Salman Beigi\\MIT
\and Andrew Drucker\\MIT
\and Bill Fefferman\\University of Chicago
\and Peter Shor\\MIT}
\date{}
\maketitle

\begin{abstract}
The class $\mathsf{QMA}\left(  k\right)  $, introduced by Kobayashi et al.,
consists of all languages that can be verified using $k$ unentangled quantum
proofs. \ Many of the simplest questions about this class have remained
embarrassingly open:\ for example, can we give any evidence that $k$ quantum
proofs are more powerful than one? \ Does $\mathsf{QMA}\left(  k\right)
=\mathsf{QMA}\left(  2\right)  $\ for $k\geq2$? \ Can $\mathsf{QMA}\left(
k\right)  $\ protocols be amplified to exponentially small error?

In this paper, we make progress on all of the above questions.

\begin{itemize}
\item We give a protocol by which a verifier can be convinced that a
\textsc{3Sat} formula\ of size $m$ is satisfiable, with constant soundness,
given $\widetilde{O}\left(  \sqrt{m}\right)  $ unentangled quantum witnesses
with $O\left(  \log m\right)  $\ qubits each. \ Our protocol relies on the
existence of very short PCPs.

\item We show that assuming a weak version of the Additivity Conjecture from
quantum information theory, any $\mathsf{QMA}\left(  2\right)  $\ protocol can
be amplified to exponentially small error, and $\mathsf{QMA}\left(  k\right)
=\mathsf{QMA}\left(  2\right)  $\ for all $k\geq2$.

\item We prove the nonexistence of \textquotedblleft perfect
disentanglers\textquotedblright\ for simulating multiple Merlins with one.

\end{itemize}
\end{abstract}

\section{Introduction\label{INTRO}}

Quantum entanglement is often described as a complicated, hard-to-understand
resource. \ But ironically, many questions in quantum computing are easiest to
answer assuming unlimited entanglement, and become much more difficult if
entanglement is \textit{not} allowed! \ One way to understand this is that
Hilbert space---the space of \textit{all} quantum states---has extremely
useful linear-algebraic properties, and when we restrict to the set of
separable states we lose many of those properties. \ So for example, finding a
quantum state that maximizes the probability of a given measurement outcome is
just a principal eigenvector problem, but finding a separable state that does
the same is $\mathsf{NP}$-hard \cite{bliertapp}.

These observations naturally give rise to a general question at the
intersection of computational complexity and entanglement theory. \ Namely:
supposing we had $k$ quantum proofs,\ could we use the promise that the proofs
were unentangled to verify mathematical statements beyond what we could verify otherwise?

\subsection{Background and Related Work\label{RELATED}}

The class $\mathsf{QMA}$, or Quantum Merlin-Arthur, consists of all languages
that admit a proof protocol in which Merlin sends Arthur a polynomial-size
quantum state $\left\vert \psi\right\rangle $, and then Arthur decides whether
to accept or reject in quantum polynomial time. \ This class was introduced by
Knill \cite{knill}, Kitaev \cite{ksv}, and Watrous \cite{watrous} as a quantum
analogue of $\mathsf{NP}$. \ By now we know a reasonable amount about
$\mathsf{QMA}$: for example, it allows amplification of success probabilities,
is contained in $\mathsf{PP}$, and has natural complete promise problems.
\ (See Aharonov and Naveh \cite{an}\ for a survey.)

In 2003, Kobayashi, Matsumoto, and Yamakami \cite{kmy}\ defined a
generalization of $\mathsf{QMA}$ called $\mathsf{QMA}\left(  k\right)  $.
\ Here there are $k$ Merlins, who send Arthur $k$ quantum proofs $\left\vert
\psi_{1}\right\rangle ,\ldots,\left\vert \psi_{k}\right\rangle $\ respectively
that are guaranteed to be unentangled with each other. \ (Thus $\mathsf{QMA}%
\left(  1\right)  =\mathsf{QMA}$.) \ Notice that in the classical case, this
generalization is completely uninteresting: we have $\mathsf{MA}\left(
k\right)  =\mathsf{MA}$\ for all $k$, since we can always simulate
$k$\ Merlins by a single Merlin who sends Arthur a concatenation of the
$k$\ proofs. \ In the quantum case, however, a single Merlin could cheat by
\textit{entangling} the $k$ proofs, and we know of no general way to detect
such entanglement.

When we try to understand $\mathsf{QMA}\left(  k\right)  $, we encounter at
least three basic questions. \ First, do multiple quantum proofs ever actually
help? \ That is, can we find some sort of evidence that $\mathsf{QMA}\left(
k\right)  \neq\mathsf{QMA}\left(  1\right)  $ for some $k$? \ Second, can
$\mathsf{QMA}\left(  k\right)  $\ protocols be amplified to exponentially
small error? \ Third, are two Merlins the most we ever need? \ That is,
does\ $\mathsf{QMA}\left(  k\right)  =\mathsf{QMA}\left(  2\right)  $ for all
$k\geq2$?\footnote{The second and third questions are motivated, in part, by
an analogy to classical \textit{multi-prover interactive proof systems}%
---where the Parallel Repetition Theorem of Raz \cite{raz:prt} and the
$\mathsf{MIP}\left(  k\right)  =\mathsf{MIP}\left(  2\right)  $\ theorem of
Ben-Or et al.\ \cite{bgkw}\ turned out to be crucial for understanding the
class $\mathsf{MIP}$.}

We know of three previous results that are relevant to the above questions.

First, in their original paper on $\mathsf{QMA}\left(  k\right)  $, Kobayashi
et al.\ \cite{kmy}\ proved that a positive answer to the second question
implies a positive answer to the third. \ That is, if $\mathsf{QMA}\left(
k\right)  $\ protocols can be amplified, then $\mathsf{QMA}\left(  k\right)
=\mathsf{QMA}\left(  2\right)  $ for all $k\geq2$.

Second, Liu, Christandl, and Verstraete \cite{lcv}\ gave a natural problem
from quantum chemistry, called \textit{pure state }$N$%
\textit{-representability}, which is in $\mathsf{QMA}\left(  2\right)  $ but
is not known to be in $\mathsf{QMA}$.

Third, Blier and Tapp \cite{bliertapp} recently (and independently of us) gave
an interesting $\mathsf{QMA}\left(  2\right)  $\ protocol for an $\mathsf{NP}%
$-complete problem, namely $3$\textsc{-Coloring}. \ In this protocol, Arthur
verifies that an $n$-vertex graph $G$ is $3$-colorable, using two unentangled
witnesses with only $O\left(  \log n\right)  $\ qubits each. \ There is a
crucial caveat, though: if $G$ is \textit{not} $3$-colorable, then Arthur can
only detect this with\ probability $\Omega\left(  1/n^{6}\right)  $ rather
than constant probability.\footnote{Indeed, if the soundness gap were constant
rather than $1/\operatorname*{poly}\left(  n\right)  $, then Blier and Tapp's
protocol could presumably be \textquotedblleft scaled up by an
exponential\textquotedblright\ to show $\mathsf{QMA}\left(  2\right)
=\mathsf{NEXP}$!}

\subsection{Our Results\label{OUR}}

In this paper, we present new results about all three problems listed
previously. \ Our main results are as follows:

\textbf{Proving }\textsc{3Sat}\textbf{ With }$\widetilde{O}\left(  \sqrt
{m}\right)  $\textbf{ Qubits.} \ In Section \ref{3SAT}, we give a protocol by
which Arthur can verify that a \textsc{3Sat}\ instance of size $m$ has a
satisfying assignment, using $O\left(  \sqrt{m}\operatorname*{polylog}%
m\right)  $ unentangled witnesses with $O\left(  \log m\right)  $\ qubits
each. \ Of course, this is a larger number of qubits than in the protocol of
Blier and Tapp \cite{bliertapp},\ but the point is that Arthur can detect
cheating with \textit{constant} probability. \ Our protocol relies on the PCP
Theorem, and in particular, on the existence of PCP's of size $O\left(
m\operatorname*{polylog}m\right)  $, which was recently shown by Dinur
\cite{dinur}.

\textbf{Weak Additivity Implies Amplification.} \ In Section \ref{AMP},\ we
reduce several open problems about $\mathsf{QMA}\left(  k\right)  $\ to weak
versions of the Additivity Conjecture\ in quantum information theory. \ (In an
earlier version of this paper, we pointed out that the weak versions would
suffice, but based our results on the original Additivity Conjecture, which
was widely believed at the time. \ Then, as the paper was undergoing final
revisions, Hastings \cite{hastings} announced a spectacular disproof of the
Additivity Conjecture.) \ In particular, we show that weak versions of
Additivity Conjecture imply that any$\ \mathsf{QMA}\left(  2\right)
$\ protocol can be amplified to exponentially small error, that the
\textquotedblleft$\mathsf{QMA}\left(  k\right)  $\ hierarchy\textquotedblright%
\ collapses to $\mathsf{QMA}\left(  2\right)  $,\ and that forcing the
Merlins' witnesses to be identical does not change the power of\ $\mathsf{QMA}%
\left(  2\right)  $.

\textbf{Nonexistence of Perfect Disentanglers.} \ In Section \ref{GDM}, we
rule out one possible approach to showing $\mathsf{QMA}\left(  2\right)
=\mathsf{QMA}$, by proving an extremely simple result\ that nevertheless seems
new and might be of interest. \ Namely, given finite-dimensional Hilbert
spaces $\mathcal{H},\mathcal{K}$, there is no quantum operation mapping the
set of all states in $\mathcal{H}$ to the set of all separable states
in$\ \mathcal{K}\otimes\mathcal{K}$.

\textit{Note:} In an earlier version of this paper, we claimed to
give evidence that $\mathsf{QMA}\left(  2\right)
\subseteq\mathsf{PSPACE}$, by showing that this would follow from
what we called the \textquotedblleft Strong Amplification
Conjecture\textquotedblright: that it is possible to amplify any
$\mathsf{QMA}\left(  2\right)  $\ protocol, in such a way that one
of the Merlin's Hilbert space dimensions remains small compared to
the inverse of the error bound. \ (The trivial upper bound is
$\mathsf{QMA}\left( 2\right)  \subseteq\mathsf{NEXP}$, which follows
by just guessing exponential-size classical descriptions of the $k$
quantum proofs.) \ However, Fernando Brandao subsequently pointed
out to us that the Strong Amplification Conjecture would actually
imply $\mathsf{QMA}\left(  2\right)
=\mathsf{QMA}$!\footnote{\label{SCHMIDT}Here is a sketch of the
argument, which we are grateful to Brandao for allowing us to
include. \ If we have two Merlins $A$ and $B$, and the witness of
$A$ has only $s\left(  n\right)  $ qubits, then any bipartite pure
state
$\left\vert \psi_{AB}\right\rangle $ can be decomposed in Schmidt form as%
\[
\left\vert \psi_{AB}\right\rangle =\sum_{i=1}^{2^{s\left(  n\right)  }}%
\lambda_{i}\left\vert \phi_{i}\right\rangle _{A}\left\vert \varphi
_{i}\right\rangle _{B},
\]
where the $\left\vert \phi_{i}\right\rangle _{A}\left\vert \varphi
_{i}\right\rangle _{B}$'s are all orthogonal to each other,
regardless of the size of $B$. \ Now, by assumption, every
unentangled state of the form $\left\vert \phi_{i}\right\rangle
_{A}\left\vert \varphi_{i}\right\rangle _{B}$ causes Arthur's
amplified protocol to accept with only a tiny probability---say,
less than $2^{-s\left(  n\right)  }$. \ But this means that Arthur's
acceptance probability on $\left\vert \psi_{AB}\right\rangle $\ can
be at most%
\[
2^{-s\left(  n\right)  }\sum_{i=1}^{2^{s\left(  n\right)  }} \left|
\lambda_{i} \right|%
\leq2^{-s\left(  n\right)  /2}%
\]
by Cauchy-Schwarz. \ Therefore, the amplified $\mathsf{QMA}\left(
2\right) $\ protocol is still sound even if the Merlins entangle
their witnesses, and hence the language being verified is in
$\mathsf{QMA}$.} \ Thus, the Strong Amplification Conjecture now
seems very unlikely be true, and while our result was correct, it
has been both superseded and effectively nullified in its import. \
We have not included it in the current version.

In the remainder of this introduction, we give some intuition behind each of
these results.

\subsection{\label{3SATINT}Proving 3SAT With $\widetilde{O}\left(  \sqrt
{m}\right)  $ Qubits}

Let $\varphi$\ be a \textsc{3Sat}\ instance with $n$ variables.\ \ Then how
long a proof does Merlin need to send Arthur, to convince him that $\varphi
$\ is satisfiable? \ (As usual, Merlin is an omniscient prover and Arthur is a
skeptical $\mathsf{BPP}$\ verifier.)

Intuitively, it seems the answer should be about $n$ bits. \ Certainly, if
sublinear-size proofs of satisfiability existed, then \textsc{3Sat}\ would be
in solvable in $2^{o\left(  n\right)  }$ time, since Arthur could just loop
over all possible proofs until he found one that worked. \ Even in the quantum
case, one can make a similar statement:\ if \textit{quantum} proofs of
satisfiability\ with $o\left(  n\right)  $ qubits\ existed, then
\textsc{3Sat}\ would have a\ $2^{o\left(  n\right)  }$-time quantum algorithm.

On the other hand, suppose Arthur is given \textit{several} quantum proofs,
which are guaranteed to be unentangled with each other. \ Then the previous
argument no longer seems to work. \ And this at least raises the possibility
that \textsc{3Sat}\ might have sublinear proofs in this setting.

We will show that this possibility is realized:

\begin{theorem}
\label{3satthm}Let $\varphi$\ be a satisfiable \textsc{3Sat} instance with $n$
variables and $m\geq n$ clauses. \ Then one can prove the satisfiability of
$\varphi$, with perfect completeness and constant soundness, using $O\left(
\sqrt{m}\operatorname*{polylog}m\right)  $ unentangled quantum proofs, each
with $O\left(  \log m\right)  $\ qubits.
\end{theorem}

In particular, if $m=O\left(  n\right)  $,\footnote{Note that setting
$m=O\left(  n\right)  $\ is fairly common in the study of \textsc{3Sat}, and
indeed, the \textquotedblleft hardest\textquotedblright\ random \textsc{3Sat}%
\ instances\ are believed to occur around $m\approx4.25n$.} then we get an
almost-quadratic improvement over the best known witness size in the classical
world\ (or for that matter, in the quantum world with one prover).

Obviously, Theorem \ref{3satthm}\ does not immediately generalize to
\textit{all} $\mathsf{NP}$-complete problems, since the blowup in reducing to
\textsc{3Sat}\ could be more than quadratic. \ It is an interesting question
for which $\mathsf{NP}$-complete problems\ an analogue of Theorem
\ref{3satthm}\ holds and for which it does not.

We now explain the intuition behind Theorem \ref{3satthm}. \ The first step in
our protocol is to reduce \textsc{3Sat} to a more convenient problem called
\textsc{2-Out-Of-4-SAT},\ where every clause has exactly four literals, and is
satisfied if and only if exactly two of the literals are. \ We also want our
\textsc{2-Out-Of-4-SAT}\ instance to be a PCP: that is, either it should be
satisfiable, or else at most a $1-\varepsilon$ fraction of clauses should be
satisfiable for some constant $\varepsilon>0$. \ Finally we want the instance
to be \textit{balanced}, meaning that every variable occurs in at most a
constant number of clauses. \ Fortunately, we can get all of this via known
classical reductions, including the \textquotedblleft tight\textquotedblright%
\ PCP Theorem of Dinur \cite{dinur}, which increase the number of variables
and clauses by at most a $\operatorname*{polylog}m$\ factor.

So suppose Arthur has applied these reductions, to obtain a balanced
\textsc{2-Out-Of-4-SAT}\ PCP instance $\phi$ with $N$ variables. \ And now
suppose Merlin sends Arthur a $\log N$-qubit\ quantum state of the form%
\[
\left\vert \psi\right\rangle =\frac{1}{\sqrt{N}}\sum_{i=1}^{N}\left(
-1\right)  ^{x_{i}}\left\vert i\right\rangle ,
\]
where $x_{1},\ldots,x_{N}\in\left\{  0,1\right\}  ^{N}$\ is the claimed
satisfying assignment for $\phi$. \ (We call a state having the above form a
\textit{proper} state.) \ Then we show that Arthur can check the veracity of
$x_{1},\ldots,x_{N}$ with perfect completeness and constant soundness. \ To do
so, Arthur simply measures $\left\vert \psi\right\rangle $\ in a basis
corresponding to the clauses of $\phi$. \ With constant probability, he will
get an outcome of the form%
\[
\left(  -1\right)  ^{x_{i}}\left\vert i\right\rangle +\left(  -1\right)
^{x_{j}}\left\vert j\right\rangle +\left(  -1\right)  ^{x_{k}}\left\vert
k\right\rangle +\left(  -1\right)  ^{x_{\ell}}\left\vert \ell\right\rangle
\]
where $\left(  i,j,k,\ell\right)  $\ is a randomly chosen clause of $\phi$.
\ Assuming this occurs, Arthur can perform a measurement that accepts with
certainty if $x_{i}+x_{j}+x_{k}+x_{\ell}=2$ and rejects with constant
probability otherwise.

Thus, if only Arthur could somehow assume $\left\vert \psi\right\rangle $\ was
proper, we would have a $\log N$-qubit witness for \textsc{3Sat}! \ The
problem, of course, is that Arthur has no way of knowing whether Merlin has
cheated and given him an improper state. \ For example, what if Merlin
concentrates the amplitude of $\left\vert \psi\right\rangle $ on some small
subset of basis states, and simply omits the other basis states?

Our key technical contribution is to show that, if Arthur gets not one but
$O(\sqrt{N})$\ copies of $\left\vert \psi\right\rangle $,\ then he can check
with constant soundness whether $\left\vert \psi\right\rangle $\ is proper or
far from any proper state. \ Indeed, even if Arthur is given $K=O(\sqrt{N}%
)$\ states $\left\vert \varphi_{1}\right\rangle ,\ldots,\left\vert \varphi
_{K}\right\rangle $\ which are not necessarily identical, so long as the
states are not entangled with each other Arthur can check with constant
soundness whether most of them are close to some proper state $\left\vert
\psi\right\rangle $. \ This then yields a protocol for \textsc{3Sat} with
constant soundness and $O(\sqrt{N})$\ unentangled proofs of size $O\left(
\log N\right)  $---for Arthur can just choose randomly whether to perform the
satisfiability test described above, or to check whether most of the
$\left\vert \varphi_{k}\right\rangle $'s are close to some proper state
$\left\vert \psi\right\rangle $.

To check that most of the states are at least close to \textit{each other},
Arthur simply has to perform a \textquotedblleft swap test\textquotedblright%
\ between (say) $\left\vert \varphi_{1}\right\rangle $\ and a random other
state $\left\vert \varphi_{k}\right\rangle $. \ So the problem is reduced to
the following: assuming most of the $\left\vert \varphi_{k}\right\rangle $'s
are close to $\left\vert \varphi_{1}\right\rangle $, how can Arthur decide
whether $\left\vert \varphi_{1}\right\rangle $\ is proper or far from any
proper state?

In our protocol, Arthur does this by first choosing a matching $\mathcal{M}%
$\ on the set $\left\{  1,\ldots,N\right\}  $\ uniformly at random. \ He then
measures each state $\left\vert \varphi_{k}\right\rangle $\ in an orthonormal
basis that contains the vectors $\left\vert i\right\rangle +\left\vert
j\right\rangle $\ and $\left\vert i\right\rangle -\left\vert j\right\rangle $
for every edge $\left(  i,j\right)  \in\mathcal{M}$.

Let us think about what happens when Arthur does this. \ Since he is
performing $O(\sqrt{N})$\ measurements on almost-identical states, and since
each measurement has $N$ possible outcomes, by using a suitable generalization
of the Birthday Paradox one can prove that with $\Omega\left(  1\right)  $
probability, Arthur will find a \textit{collision}: that is, two outcomes of
the form $\left\vert i\right\rangle \pm\left\vert j\right\rangle $, for the
same edge $\left(  i,j\right)  \in\mathcal{M}$. \ So suppose this happens.
\ Then if the $\left\vert \varphi_{k}\right\rangle $'s are all equal to a
proper state $\left\vert \psi\right\rangle =\sum_{i=1}^{N}\left(  -1\right)
^{x_{i}}\left\vert i\right\rangle $, the two outcomes will clearly
\textquotedblleft agree\textquotedblright: that is, they will either both be
$\left\vert i\right\rangle +\left\vert j\right\rangle $\ (if $x_{i}=x_{j}$) or
both be $\left\vert i\right\rangle -\left\vert j\right\rangle $ (if $x_{i}\neq
x_{j}$). \ On the other hand, suppose the\ $\left\vert \varphi_{k}%
\right\rangle $'s are far from any proper state. \ In that case, we show that
the outcomes will \textquotedblleft disagree\textquotedblright\ (that is, one
will be $\left\vert i\right\rangle +\left\vert j\right\rangle $\ and the other
will be $\left\vert i\right\rangle -\left\vert j\right\rangle $) with
$\Omega\left(  1\right)  $ probability.

To understand why, consider that there are two ways for a state $\left\vert
\varphi\right\rangle =\sum_{i=1}^{N}\alpha_{i}\left\vert i\right\rangle $\ to
be far from proper. \ First, the probability distribution $\left(  \left\vert
\alpha_{1}\right\vert ^{2},\ldots,\left\vert \alpha_{N}\right\vert
^{2}\right)  $, which corresponds to measuring $\left\vert \varphi
\right\rangle $\ in the standard basis, could be far from the uniform
distribution. \ Second, the $\alpha_{i}$'s could be roughly equal in
magnitude, but they could have complex phases that cause $\left\vert
\varphi\right\rangle $\ to be far from any state involving positive and
negative real amplitudes only. \ In either case, though, if Arthur measures
according to a random matching $\mathcal{M}$, then with high probability he
will obtain an outcome $\alpha_{i}\left\vert i\right\rangle +\alpha
_{j}\left\vert j\right\rangle $\ that is not close to either $\left\vert
i\right\rangle +\left\vert j\right\rangle $\ or $\left\vert i\right\rangle
-\left\vert j\right\rangle $.

As one would imagine, making all of these claims quantitative and proving them
requires a good deal of work.

The reason we need the assumption of unentanglement is that without it,
cheating Merlins might correlate their states in such a way that a swap test
between any two states passes with certainty, and yet no collisions are ever
observed. \ As we point out in Section \ref{REMARKS}, it seems unlikely that
the assumption of unentanglement can be removed, since this would lead to a
$2^{\widetilde{O}\left(  \sqrt{m}\right)  }$-time classical algorithm for
\textsc{3Sat}. \ On the other hand, we believe it should be possible to
improve our protocol to one involving only \textit{two} unentangled proofs.
\ This is a problem we leave to future work.

\subsection{Weak Additivity Implies Amplification\label{AMPINT}}

In the one-prover case, it is easy to amplify a $\mathsf{QMA}$\ protocol with
constant error to a protocol with exponentially small error. \ Merlin simply
sends Arthur $m=\operatorname*{poly}\left(  n\right)  $ copies of his proof;
then Arthur checks each of the copies and outputs the majority answer. \ To
show that this works, the key observation is that \textit{Merlin cannot gain
anything by entangling the }$\mathit{m}$\textit{\ proofs}. \ Indeed, because
of the convexity of Arthur's acceptance probability, Merlin might as well send
Arthur an unentangled state $\left\vert \psi\right\rangle ^{\otimes m}$, in
which case the completeness and soundness errors will decrease exponentially
with $m$ by the usual\ Chernoff bound.

Now suppose we try the same argument for $\mathsf{QMA}\left(  2\right)  $.
\ If we ask each Merlin to send $m$ copies of his state, each Merlin might
cheat by instead sending an entangled state on $m$ registers. \ And in that
case, as soon as Arthur checks the first copy (consisting of one register from
Merlin$_{A}$ and one from Merlin$_{B}$), \textit{his doing so might create
entanglement in the remaining copies where there was none before!} \ This is
because of a counterintuitive phenomenon called \textit{entanglement swapping}
\cite{zzhe}, by which two quantum systems\ that have never interacted in the
past can nevertheless become entangled, provided those systems are entangled
with \textit{other} systems on which an entangling measurement is performed.

Let us give a small illustration of this phenomenon. \ Suppose that each
\textquotedblleft proof\textquotedblright\ is a single qubit, and that Arthur
asks for two proofs from each Merlin (thus, $4$ qubits in total). \ Then\ if
Merlin$_{A}$ is dishonest, he might send Arthur the entangled state
$\left\vert \psi_{A}\right\rangle =\left\vert 00\right\rangle +\left\vert
11\right\rangle $, and likewise Merlin$_{B}$ might send Arthur $\left\vert
\psi_{B}\right\rangle =\left\vert 00\right\rangle +\left\vert 11\right\rangle
$ (omitting normalization). \ Now suppose Arthur measures the qubits
$\left\vert \psi_{A}\right\rangle _{\left(  1\right)  }$\ and $\left\vert
\psi_{B}\right\rangle _{\left(  1\right)  }$\ in the \textquotedblleft Bell
basis,\textquotedblright\ consisting of the four entangled states $\left\vert
00\right\rangle +\left\vert 11\right\rangle $, $\left\vert 00\right\rangle
-\left\vert 11\right\rangle $, $\left\vert 01\right\rangle +\left\vert
10\right\rangle $, and $\left\vert 01\right\rangle -\left\vert 10\right\rangle
$. \ Then conditioned on the outcome of this measurement, it is not hard to
see that the joint state of $\left\vert \psi_{A}\right\rangle _{\left(
2\right)  }$\ and $\left\vert \psi_{B}\right\rangle _{\left(  2\right)  }%
$\ will also be entangled.\footnote{Indeed, this example can be seen as a
special case of \textit{quantum teleportation} \cite{bbcjpw}: Arthur uses the
entanglement between Merlin$_{A}$'s left and right registers, as well as
between Merlin$_{B}$'s left and right registers, to teleport an entangled
state into the two right registers by acting only on the two left registers.}

Of course, as soon as the remaining $m-1$\ copies become entangled, we lose
our soundness guarantee and the proof of amplification fails.

Nevertheless, there is a natural amplification procedure that seems like it
\textit{ought} to be robust against such \textquotedblleft
pathological\textquotedblright\ behavior. \ Suppose Arthur chooses the number
of copies $m$ to be very large, say $n^{10}$ (much larger than the number of
copies he is actually going to check), and suppose that each copy he
\textit{does} check is chosen uniformly at random. \ Then whatever
entanglement Arthur produces during the checking process ought be
\textquotedblleft spread out\textquotedblright\ among the copies, so that with
high probability, every copy that Arthur actually encounters is close to separable.

It follows, from the \textquotedblleft finite quantum de Finetti
theorem\textquotedblright\ of K\"{o}nig and Renner \cite{konig}, that if the
number of copies were large enough then the above argument would work.
\ Unfortunately, the number of copies needs to be exponential in $n$ for that
theorem to apply.

We will show that the argument works with $\operatorname*{poly}\left(
n\right)  $\ copies, provided one can formalize terms like \textquotedblleft
spread out\textquotedblright\ and \textquotedblleft close to
separable\textquotedblright\ using a suitable measure of entanglement. \ The
only problem, then, is that a measure of entanglement with the properties we
want is not yet known to exist! \ Informally, we want an entanglement measure
$E$\ that

\begin{enumerate}
\item[(i)] is \textit{superadditive} (meaning it \textquotedblleft spreads
itself out\textquotedblright\ among registers), and

\item[(ii)] is \textit{faithful} (meaning if $E\left(  \rho\right)  $\ is
polynomially small then $\rho$\ is polynomially close to a separable state in
trace distance).
\end{enumerate}

Among existing entanglement measures, there is one---the \textit{entanglement
of formation} $E_{F}$, introduced by Bennett et al.\ \cite{bdsw}---that is
known to satisfy (ii), and that until recently was conjectured to satisfy
(i).\footnote{There is also another measure---the \textit{squashed
entanglement} $E_{sq}$, introduced by Christandl and Winter \cite{squashed}%
---that is known to satisfy (i), but unfortunately can be shown \textit{not}
to satisfy (ii).} \ This conjecture is known to be equivalent to the
Additivity Conjecture from quantum information theory \cite{shor:add}. \ Thus,
an earlier version of this paper assumed the Additivity Conjecture in proving
several results.

In a dramatic recent development, Hastings \cite{hastings}\ has shown that the
Additivity Conjecture is false. \ Fortunately, our results require only weak,
asymptotic versions of the Additivity Conjecture, which we still conjecture
are true, and which are stated formally in Section \ref{EF}.

Our first result says that, if a weak Additivity Conjecture holds,
then any $\mathsf{QMA}\left(  2\right)  $ protocol\ can be amplified
to exponentially small error. \ We also prove, unconditionally, that
any $\mathsf{QMA}\left( k\right)  $\ protocol with constant
soundness can be simulated by a $\mathsf{QMA}\left(  2\right)  $
protocol\ with $\Omega\left(  1/k\right) $\ soundness. \ Combining
these two results, we find that if a weak Additivity Conjecture
holds, then $\mathsf{QMA}\left(  k\right)  =\mathsf{QMA}\left(
2\right)  $\ for all $k\geq2$.\footnote{Kobayashi et al. (personal
communication) have shown independently of us that if
$\mathsf{QMA}\left( 2\right)  $\ protocols can be amplified to
exponentially small error, then $\mathsf{QMA}\left(  k\right)
=\mathsf{QMA}\left(  2\right)  $\ for all $k\geq2$.} \ Another
interesting consequence we get is that, assuming a weak Additivity
Conjecture, $k$ Merlins who all send copies of the same witness
yield the same computational power as $k$ Merlins who can send
different witnesses.

\subsection{\label{GDMINT}Nonexistence of Perfect Disentanglers}

While we now have a few examples where multiple quantum proofs seem to
help---such as the \textsc{3Sat} protocol of this paper, and the pure state
$N$-representability problem \cite{lcv}---we still have no \textquotedblleft
complexity-theoretic\textquotedblright\ evidence that $\mathsf{QMA}\left(
2\right)  \neq\mathsf{QMA}$. \ Indeed, even proving an oracle separation
between $\mathsf{QMA}\left(  2\right)  $ and $\mathsf{QMA}$\ seems extremely difficult.

Thus, let us consider the other direction and try to prove these classes are
the same. \ Potentially the first approach would be to equip Arthur with a
\textit{disentangler}:\ that is, a quantum operation that would convert
Merlin's (possibly-entangled) witness into a separable witness, and thereby
let Arthur simulate a $\mathsf{QMA}\left(  2\right)  $\ protocol in
$\mathsf{QMA}$. \ In this paper we take a first step in the study of
disentanglers, by proving that in finite-dimensional Hilbert spaces, there is
no operation that produces all and only the separable states as
output\textit{.}

Note that, if we are willing to settle for there being an output
\textit{close} to every separable state, then a disentangler does exist:
simply take as input a classical description of the separable state $\sigma
$\ to be prepared, measure that description in the computational basis, and
then prepare $\sigma$.\footnote{This example also shows that our result fails
if the input space is infinite-dimensional---for then one could give an
infinitely-precise description of $\sigma$.}

Likewise, if we are willing to settle for every output being close to a
separable state, then a disentangler also exists. \ For some sufficiently
large $N$, take as input a quantum state on registers $R_{0},R_{1}%
,\ldots,R_{N}$, choose an index $i\in\left[  N\right]  $ uniformly at random,
and output the joint state of $R_{0}$\ and $R_{i}$\ (discarding everything
else). \ It follows, from the finite quantum de Finetti theorem \cite{konig},
that with high probability this state will be close to separable.

The key problem with both of these approaches is that \textit{the input
Hilbert space needs to be exponentially larger than the output Hilbert space}.
\ In the case of the \textquotedblleft de Finetti approach,\textquotedblright%
\ this follows from considering the maximally antisymmetric state%
\[
\frac{1}{\sqrt{N!}}\sum_{\sigma\in S_{N}}\left(  -1\right)
^{\operatorname*{sgn}\left(  \sigma\right)  }\left\vert \sigma\left(
1\right)  \right\rangle \cdots\left\vert \sigma\left(  N\right)  \right\rangle
,
\]
which has the properties that

\begin{enumerate}
\item[(i)] there are exponentially many registers (as a function of $n=\log
N$, the size of a given register), but

\item[(ii)] the reduced state of any two registers is far from a separable state.
\end{enumerate}

Watrous (personal communication) has conjectured that this exponentiality is
an unavoidable feature of any approximate disentangler.\ \ Proving or
disproving this remains one of the central open problems about $\mathsf{QMA}%
\left(  2\right)  $.

\subsection{Table of Contents\label{ORG}}

The rest of the paper is organized as follows.\bigskip%

\begin{tabular}
[c]{ll}%
Section \ref{PRELIM} & Preliminaries\\
Section \ref{3SAT} & Proving \textsc{3Sat}\ with $\widetilde{O}\left(
\sqrt{m}\right)  $ qubits\\
Section \ref{AMP} & Weak additivity implies amplification (as well as
$\mathsf{QMA}\left(  k\right)  =\mathsf{QMA}\left(  2\right)  $, etc.)\\
Section \ref{GDM} & Nonexistence of perfect disentanglers\\
Section \ref{OPEN} & List of open problems
\end{tabular}
\bigskip

\section{Preliminaries\label{PRELIM}}

In this section, we first define the complexity class $\mathsf{QMA}\left(
k,a,b\right)  $, or Quantum Merlin-Arthur with $k$ unentangled witnesses and
error bounds $a,b$, and state some basic facts and conjectures about this
class. \ We then survey some concepts from quantum information theory we will
need, including trace distance, superoperators, and the swap test.

\subsection{\label{MQMA}Multiple-Prover $\mathsf{QMA}$}

\begin{definition}
A language $L$ is in $\mathsf{QMA}\left(  k,a,b\right)  $\ if there exists a
polynomial-time quantum algorithm $Q$\ such that for all inputs $x\in\left\{
0,1\right\}  ^{n}$:

\begin{enumerate}
\item[(i)] If $x\in L$\ then there exist witnesses $\left\vert \psi
_{1}\right\rangle ,\ldots,\left\vert \psi_{k}\right\rangle $, with
$\operatorname*{poly}\left(  n\right)  $\ qubits each, such that $Q$\ accepts
with probability at least $b$\ given $\left\vert x\right\rangle \otimes
\left\vert \psi_{1}\right\rangle \otimes\cdots\otimes\left\vert \psi
_{k}\right\rangle $.

\item[(ii)] If $x\notin L$\ then $Q$\ accepts with probability at most
$a$\ given $\left\vert x\right\rangle \otimes\left\vert \psi_{1}\right\rangle
\otimes\cdots\otimes\left\vert \psi_{k}\right\rangle $, for all $\left\vert
\psi_{1}\right\rangle ,\ldots,\left\vert \psi_{k}\right\rangle $.
\end{enumerate}

As a convention, we also define $\mathsf{QMA}\left(  k\right)  :=\mathsf{QMA}%
\left(  k,1/3,2/3\right)  $, and $\mathsf{QMA}:=\mathsf{QMA}\left(  1\right)
$.\footnote{For purposes of definition, we assume we have fixed a specific
machine model (e.g., a universal set of quantum gates)---though if the
Amplification Conjecture to be discussed shortly holds, then this choice will
turn out not to matter.}
\end{definition}

The above definition makes sense for all integers $k$\ from $1$ up to
$\operatorname*{poly}\left(  n\right)  $, and nonnegative real functions
$2^{-\operatorname*{poly}\left(  n\right)  }\leq a\left(  n\right)  <b\left(
n\right)  \leq1-2^{-\operatorname*{poly}\left(  n\right)  }$.

In the one-prover case, we know\ that $\mathsf{QMA}\left(  1,1/3,2/3\right)
=\mathsf{QMA}\left(  1,2^{-p\left(  n\right)  },1-2^{-p\left(  n\right)
}\right)  $ for all polynomials $p$ (see \cite{mw} for example). \ This is
what justifies the convention $\mathsf{QMA}\left(  1\right)  =\mathsf{QMA}%
\left(  1,1/3,2/3\right)  $. \ By contrast, we do not yet know whether the
convention $\mathsf{QMA}\left(  k\right)  =\mathsf{QMA}\left(
k,1/3,2/3\right)  $\ is justified for $k\geq2$.\ \ That it \textit{is}
justified is the content of the following conjecture:

\begin{conjecture}
[Amplification]$\mathsf{QMA}\left(  k,a,b\right)  =\mathsf{QMA}\left(
k,2^{-p\left(  n\right)  },1-2^{-p\left(  n\right)  }\right)  $ for all $k$,
all $a<b$\ with $b-a=\Omega\left(  1/\operatorname*{poly}\left(  n\right)
\right)  $, and all polynomials $p$.\label{ampconj}
\end{conjecture}

One is tempted to make an even stronger conjecture: that the entire hierarchy
of $\mathsf{QMA}\left(  k,a,b\right)  $'s we have defined collapses to just
two complexity classes, namely $\mathsf{QMA}$\ and $\mathsf{QMA}\left(
2\right)  $.

\begin{conjecture}
[Collapse]$\mathsf{QMA}\left(  k,a,b\right)  =\mathsf{QMA}\left(
2,2^{-p\left(  n\right)  },1-2^{-p\left(  n\right)  }\right)  $ for all
$k\geq2$, all $a<b$\ with $b-a=\Omega\left(  1/\operatorname*{poly}\left(
n\right)  \right)  $, and all polynomials $p$.\label{collconj}
\end{conjecture}

The main progress so far on these conjectures has been due to Kobayashi et al.
\cite{kmy},\ who showed that the Amplification and Collapse Conjectures are
actually equivalent:

\begin{theorem}
[\textbf{\cite{kmy}}]\label{kmythm}Conjecture \ref{ampconj}\ implies
Conjecture \ref{collconj}.
\end{theorem}

Let us observe that one can make the \textit{completeness} error (though not
the soundness error) exponentially small, using a simple argument based on
Markov's inequality. \ We will need this observation in Section \ref{AMP}.

\begin{lemma}
\label{onesidedamp}$\mathsf{QMA}\left(  k,a,b\right)  \subseteq\mathsf{QMA}%
\left(  k,1-\left(  b-a\right)  ,1-2^{-p\left(  n\right)  }\right)  $ for all
$k$, all $a<b<1$, and all polynomials $p$.
\end{lemma}

\begin{proof}
We use the following protocol. \ Each Merlin provides
$m=C\cdot\frac{p\left( n\right)  }{\left(  b-a\right)  ^{2}}$\
registers for some constant $C$.\ \ Then Arthur runs his
verification procedure $m$ times in parallel, once with each
$k$-tuple\ of registers, and accepts if and only if at least a $d$\
fraction of invocations accept, for some $d$ slightly less than
$b$.

To show completeness, we use a Chernoff bound. \ Assuming the
Merlins are honest, each one simply provides $m$ copies of his
witness. \ Then on each invocation, Arthur accepts with independent
probability at least $b$. \ So assuming we chose a sufficiently
large constant $C$, the probability that Arthur accepts less than
$dm$\ times is at most $2^{-p\left(  n\right)  }$.

To show soundness, we use Markov's inequality. \ The expected number
of accepting invocations is at most $am$ (by linearity of
expectation, this is true even if the registers are entangled). \
Hence the probability that this number exceeds $dm$\ is at most
$a/d$, which we can ensure is less than $1-\left( b-a\right)  $\ by
choosing $d\in\left( \frac{a}{1-\left(  b-a\right) },b\right)  $\
(note that such a $d$ must exist by the assumption $a<b<1$).
\end{proof}

\subsection{Quantum Information\label{QI}}

We now review some quantum information concepts that we will need. \ For more
details see Nielsen and Chuang \cite{nc} for example.

Given two mixed states $\rho$\ and $\sigma$, their \textit{trace distance} is
$\left\Vert \rho-\sigma\right\Vert _{\operatorname*{tr}}:=\frac{1}{2}%
\sum_{i=1}^{n}\left\vert \lambda_{i}\right\vert $, where $\left(  \lambda
_{1},\ldots,\lambda_{n}\right)  $\ are the eigenvalues of $\rho-\sigma$. \ We
say $\sigma$\ is $\varepsilon$\textit{-close} to $\rho$\ if $\left\Vert
\rho-\sigma\right\Vert _{\operatorname*{tr}}\leq\varepsilon$, and
$\varepsilon$\textit{-far}\ otherwise. \ The importance of trace distance
comes from the following fact:

\begin{proposition}
\label{closeprop}Suppose $\sigma$\ is\ $\varepsilon$-close to $\rho$. \ Then
any measurement that accepts $\rho$\ with probability $p$, accepts $\sigma
$\ with probability at most $p+\varepsilon$.
\end{proposition}

Trace distance also satisfies the triangle inequality:

\begin{proposition}
\label{triangle}$\left\Vert \rho-\sigma\right\Vert _{\operatorname*{tr}%
}+\left\Vert \sigma-\xi\right\Vert _{\operatorname*{tr}}\geq\left\Vert
\rho-\xi\right\Vert _{\operatorname*{tr}}$ for all $\rho,\sigma,\xi$.
\end{proposition}

Given a pure state $\left\vert \psi\right\rangle $\ and a mixed state $\rho$,
their \textit{squared fidelity} $\left\langle \psi|\rho|\psi\right\rangle $ is
the probability of obtaining $\left\vert \psi\right\rangle $\ as the result of
a projective measurement on $\rho$. \ Squared fidelity behaves nicely under
tensor products:

\begin{proposition}
\label{closetopure}Given a $k$-partite state $\rho^{A_{1}A_{2}\cdots A_{k}}$,
suppose there are pure states $\left\vert \psi_{1}\right\rangle ,\ldots
,\left\vert \psi_{k}\right\rangle $ such that $\left\langle \psi_{i}%
|\rho^{A_{i}}|\psi_{i}\right\rangle \geq1-\varepsilon_{i}$ for all $i$. \ Let
$\left\vert \Psi\right\rangle :=\left\vert \psi_{1}\right\rangle \otimes
\cdots\otimes\left\vert \psi_{k}\right\rangle $ and $\varepsilon
:=\varepsilon_{1}+\cdots+\varepsilon_{k}$. \ Then $\left\langle \Psi
|\rho^{A_{1}A_{2}\cdots A_{k}}|\Psi\right\rangle \geq1-\varepsilon$.
\end{proposition}

\begin{proof}
We can assume without loss of generality that $\left\vert \psi_{i}%
\right\rangle =\left\vert 0\right\rangle $ for all $i$. \ Then each
$\rho^{A_{i}}$, when measured in the standard basis, yields the outcome
$\left\vert 0\right\rangle $\ with probability at least $1-\varepsilon_{i}$.
\ By the union bound, it follows that $\rho^{A_{1}A_{2}\cdots A_{k}}$, when
measured in the standard basis, yields the outcome $\left\vert \Psi
\right\rangle =\left\vert 0\right\rangle ^{\otimes k}$\ with probability at
least $1-\varepsilon$. \ Hence $\left\langle \Psi|\rho^{A_{1}A_{2}\cdots
A_{k}}|\Psi\right\rangle \geq1-\varepsilon$.
\end{proof}

Trace distance and squared fidelity are related to each other as follows:

\begin{proposition}
\label{fidelityprop}$\left\langle \psi|\rho|\psi\right\rangle +\left\Vert
\rho-\left\vert \psi\right\rangle \left\langle \psi\right\vert \right\Vert
_{\operatorname*{tr}}^{2}\leq1$ for all $\rho$\ and $\left\vert \psi
\right\rangle $.
\end{proposition}

The most general kind of operation on quantum states is called a
\textit{superoperator}. \ Any superoperator $\Phi$\ acting on $n$ qubits can
be expressed in the following \textit{operator-sum representation}:
$\Phi\left(  \rho\right)  =\sum_{i=1}^{2^{2n}}E_{i}\rho E_{i}^{\dagger}$,
where $\sum_{i=1}^{2^{2n}}E_{i}^{\dagger}E_{i}=I$.

Given a product state $\rho\otimes\sigma$, the \textit{swap test} is a quantum
operation that measures the overlap between $\rho$\ and $\sigma$. \ The test
accepts with probability $\frac{1+\operatorname*{tr}\left(  \rho\sigma\right)
}{2}$ and rejects otherwise. \ The swap test can also reveal information about
the purity of a state, as follows:

\begin{proposition}
\label{swaptest}Suppose $\left\langle \psi|\rho|\psi\right\rangle
<1-\varepsilon$ for all pure states $\left\vert \psi\right\rangle $. \ Then a
swap test between $\rho$\ and any other state rejects with probability greater
than $\varepsilon/2$.
\end{proposition}

\begin{proof}
Choose a basis that diagonalizes $\rho$, so that $\rho=\operatorname*{diag}%
\left(  \lambda_{1},\ldots,\lambda_{N}\right)  $ where $\lambda_{1}%
,\ldots,\lambda_{N}$\ are $\rho$'s eigenvalues. \ By assumption, $\lambda
_{i}<1-\varepsilon$\ for every $i$. \ So given any mixed state $\sigma$, a
swap test between $\rho$\ and $\sigma$ accepts with probability%
\[
\frac{1+\operatorname*{tr}\left(  \rho\sigma\right)  }{2}=\frac{1}{2}+\frac
{1}{2}\sum_{i=1}^{N}\lambda_{i}\sigma_{ii}<\frac{1}{2}+\frac{1-\varepsilon}%
{2}\sum_{i=1}^{N}\sigma_{ii}=1-\frac{\varepsilon}{2}.
\]
\end{proof}

\section{Proving 3SAT With Quadratically Fewer Qubits\label{3SAT}}

We now present our protocol for proving the satisfiability of a \textsc{3Sat}
instance of size $m$, using $\widetilde{O}\left(  \sqrt{m}\right)
$\ unentangled quantum proofs with $O\left(  \log m\right)  $\ qubits each.
\ For ease of presentation, the protocol will be broken into a sequence of
four steps:

\begin{enumerate}
\item[(1)] In Section \ref{CLASSICAL}, we give a sequence of classical
reductions, from the original \textsc{3Sat}\ problem to a different
$\mathsf{NP}$-complete problem that we will actually use.

\item[(2)] In Section \ref{PROPER}, we describe a protocol for the special
case where Merlin's message to Arthur is \textquotedblleft
proper\textquotedblright: that is, of the form $\frac{1}{\sqrt{N}}\sum
_{i=1}^{N}\left(  -1\right)  ^{x_{i}}\left\vert i\right\rangle $\ for some
Boolean $x_{1},\ldots,x_{N}$.

\item[(3)] In Section \ref{SYMMETRIC}, we generalize our protocol to the case
where the Merlins send Arthur $\widetilde{O}\left(  \sqrt{m}\right)  $
witnesses, which are not necessarily proper but which are guaranteed to be
identical to each other.

\item[(4)] In Section \ref{GENERAL}, we remove the restriction that the states
be identical.
\end{enumerate}

We end in Section \ref{REMARKS}\ with some general observations about our
protocol and the prospects for improving it further.

\subsection{Classical Reductions\label{CLASSICAL}}

It will be convenient to work not with \textsc{3Sat} but with a related
problem called \textsc{2-Out-Of-4-SAT},\ in which every clause has exactly
four literals, and is satisfied if and only if exactly two of the literals
are. \ We will also need our \textsc{2-Out-Of-4-SAT}\ instance to be a PCP,
and to be balanced (that is, every variable should appear in at most $O\left(
1\right)  $\ clauses).\ \ The following lemma shows how to get everything we
want with only a polylogarithmic blowup in the number of variables and clauses.

\begin{lemma}
\label{reducelem}There exists a polynomial-time Karp reduction that maps a
\textsc{3Sat}\ instance $\varphi$\ to a \textsc{2-Out-Of-4-SAT} instance
$\phi$, and that has the following properties:

\begin{enumerate}
\item[(i)] If $\varphi$ has $n$ variables and $m\geq n$ clauses, then $\phi$
has $O\left(  m\operatorname*{polylog}m\right)  $\ variables and $O\left(
m\operatorname*{polylog}m\right)  $\ clauses.

\item[(ii)] Every variable of $\phi$ occurs in at most $c$ clauses, for some
constant $c$.

\item[(iii)] The reduction is a PCP (meaning that satisfiable instances map to
satisfiable instances, while unsatisfiable instances map to instances that are
$\varepsilon$-far\ from satisfiable for some constant $\varepsilon>0$).
\end{enumerate}
\end{lemma}

\begin{proof}
Given a \textsc{3Sat}\ instance $\varphi$, we first amplify its soundness gap
to a constant using the celebrated method of Dinur \cite{dinur}. \ Next we use
a reduction due to Papadimitriou and Yannakakis \cite{py}, which makes every
variable occur in exactly $29$ clauses, without destroying the soundness gap.
\ Finally we use a gadget due to Khanna et al.\ \cite{kstw}, which converts
from \textsc{3Sat}\ to \textsc{2-Out-Of-4-SAT}, without destroying either the
soundness gap or the property of being balanced. \ Note that the reduction of
Dinur \cite{dinur}\ incurs only a polylogarithmic blowup in the total size of
the instance, while the other two reductions incur a constant blowup.
\end{proof}

\subsection{The Proper State Case\label{PROPER}}

Suppose Arthur has applied Lemma \ref{reducelem}, to obtain a balanced
\textsc{2-Out-Of-4-SAT}\ instance $\phi$ with $N=O\left(
m\operatorname*{polylog}m\right)  $ variables, $M=O\left(
m\operatorname*{polylog}m\right)  $\ clauses, and a constant soundness gap
$\varepsilon>0$. \ And now suppose Merlin sends Arthur a $\log N$-qubit\ state
of the form%
\[
\left\vert \psi\right\rangle =\frac{1}{\sqrt{N}}\sum_{i=1}^{N}\left(
-1\right)  ^{x_{i}}\left\vert i\right\rangle ,
\]
where $x_{1},\ldots,x_{N}\in\left\{  0,1\right\}  ^{N}$\ is a claimed
satisfying assignment for $\phi$. \ Call a state having the above form (for
some Boolean $x_{i}$'s) a \textit{proper} state. \ Then we claim the following:

\begin{lemma}
\label{properlem}Assuming $\left\vert \psi\right\rangle $ is proper, Arthur
can check whether $\phi$\ is satisfiable with perfect completeness and
constant soundness.
\end{lemma}

\begin{proof}
To perform the check, Arthur uses the following \textit{Satisfiability Test}.
\ First he partitions the clauses of $\phi$\ into a constant number of blocks
$B_{1},\ldots,B_{s}$, such that within each block, no two clauses share a
variable. \ Such a partition clearly exists by the assumption that $\phi$\ is
balanced, and furthermore can be found efficiently (e.g., using a greedy
algorithm). \ Next he chooses one of the blocks $B_{r}$\ uniformly at random,
and measures $\left\vert \psi\right\rangle $\ in an orthonormal basis with one
projector for each clause in $B_{r}$. \ Because a single block in the
partition of clauses does not necessarily cover all the variables, it is
possible that the measurement result will not correspond to any clause in
$B_{r}$, in which case Arthur accepts. \ However, suppose that the measurement
yields the following reduced state, for some random clause $C_{ijk\ell
}:=\left(  i,j,k,\ell\right)  $\ in $B_{r}$:%
\[
\left\vert \psi_{ijkl}\right\rangle :=\frac{1}{2}\left[  \left(  -1\right)
^{x_{i}}\left\vert i\right\rangle +\left(  -1\right)  ^{x_{j}}\left\vert
j\right\rangle +\left(  -1\right)  ^{x_{k}}\left\vert k\right\rangle +\left(
-1\right)  ^{x_{\ell}}\left\vert \ell\right\rangle \right]  .
\]
Notice that, of the $16$ possible assignments to the variables\
$\left( x_{i},x_{j},x_{k},x_{\ell}\right)  $, six of them satisfy
$C_{ijk\ell}$, and those six lead to three states $\left\vert
\psi_{ijk\ell}\right\rangle $ that are orthogonal to one another (as
well as the negations of those states, which are essentially the
same). \ It follows that Arthur can perform a projective measurement
on $\left\vert \psi_{ijk\ell}\right\rangle $, which accepts with
probability $1$ if $C_{ijk\ell}$\ is satisfied, and rejects with
constant probability if $C_{ijk\ell}$\ is unsatisfied. Furthermore,
because the number of blocks $B_{r}$\ is a constant, each of the $M$
clauses of $\phi$\ is checked in this test with probability
$\Omega\left(  1/M\right)  $. \ And we know that, if
$x_{1},\ldots,x_{N}$\ is \textit{not} a satisfying\ assignment for
$\phi$, then a constant fraction of the clauses will be unsatisfied.

Putting everything together, we find that if $\phi$\ is satisfiable,
then the Satisfiability Test\ accepts $\left\vert \psi\right\rangle
$\ with probability $1$; while if $\phi$\ is unsatisfiable, then it
rejects with constant probability.
\end{proof}

\subsection{The Symmetric Case\label{SYMMETRIC}}

Thus, the problem we need to solve is \textquotedblleft
merely\textquotedblright\ how to force Merlin to send a proper state. \ For
example, how can Arthur prevent a cheating Merlin from concentrating the
amplitude of $\left\vert \psi\right\rangle $ on some subset of basis states
for which the Satisfiability Test accepts, and omitting the other basis states?

To solve this problem, Arthur is going to need more Merlins. \ In particular,
let us suppose there are $K=\Theta(\sqrt{N})$ unentangled Merlins, who send
Arthur $\log N$-qubit states $\left\vert \varphi_{1}\right\rangle
,\ldots,\left\vert \varphi_{K}\right\rangle $ respectively. \ By convexity, we
can assume without loss of generality that these states are pure. \ For the
time being, we also assume that the states are identical; that is, $\left\vert
\varphi_{i}\right\rangle =\left\vert \varphi\right\rangle $\ for all
$i\in\left[  K\right]  $. \ Given these states, Arthur performs one of the
following two tests, each with probability $1/2$:

\begin{quotation}
\textbf{Satisfiability Test:} Arthur chooses any copy of $\left\vert
\varphi\right\rangle $, and performs the Satisfiability Test described in
Section \ref{PROPER}.

\textbf{Uniformity Test:} Arthur chooses a matching $\mathcal{M}$\ on $\left[
N\right]  $ uniformly at random. \ He then measures each copy of\ $\left\vert
\varphi\right\rangle $\ in an orthonormal basis, which contains the vectors%
\[
\frac{\left\vert i\right\rangle +\left\vert j\right\rangle }{\sqrt{2}}%
,\frac{\left\vert i\right\rangle -\left\vert j\right\rangle }{\sqrt{2}}%
\]
for every edge $\left(  i,j\right)  \in\mathcal{M}$. \ If for some $\left(
i,j\right)  \in\mathcal{M}$, the two outcomes $\frac{\left\vert i\right\rangle
+\left\vert j\right\rangle }{\sqrt{2}}$\ and $\frac{\left\vert i\right\rangle
-\left\vert j\right\rangle }{\sqrt{2}}$\ both occur among the $K$ measurement
outcomes, then Arthur rejects. \ Otherwise he accepts.
\end{quotation}

It is clear that the above protocol has perfect completeness. \ For if $\phi$
is satisfiable, then the Merlins can just send $K$ copies of a proper state
$\left\vert \psi\right\rangle $ corresponding to a satisfying assignment for
$\phi$. \ In that case, both tests will accept with probability $1$. \ Our
goal is to prove the following:

\begin{theorem}
\label{symthm}The protocol has constant soundness (again, assuming the
$\left\vert \varphi_{i}\right\rangle $'s\ are all identical).
\end{theorem}

To prove Theorem \ref{symthm}, we need to show that if $\phi$\ is
unsatisfiable, then one of the two tests rejects with constant probability.
\ There are two cases. \ First suppose $\left\vert \varphi\right\rangle $\ is
$\varepsilon$-close in trace distance to some proper state $\left\vert
\psi\right\rangle $. \ Then provided we choose $\varepsilon>0$\ sufficiently
small, Lemma \ref{properlem}, combined with\ Proposition \ref{closeprop}%
,\ already implies that the Satisfiability Test rejects with constant
probability. \ So our task reduces to proving the following:

\begin{claim}
\label{unifclaim}Suppose $\left\vert \varphi\right\rangle $\ is $\varepsilon
$-far in trace distance from any proper state $\left\vert \psi\right\rangle $,
for some $\varepsilon>0$. Then the Uniformity Test rejects with some constant
probability $\delta\left(  \varepsilon\right)  >0$.
\end{claim}

In analyzing the Uniformity Test, we say that Arthur \textit{finds a
collision} if he obtains two measurement outcomes of the form $\left\vert
i\right\rangle \pm\left\vert j\right\rangle $\ for the same $\left(
i,j\right)  $ pair,\ and that he \textit{finds a disagreement} if one of the
outcomes is $\left\vert i\right\rangle +\left\vert j\right\rangle $\ and the
other is $\left\vert i\right\rangle -\left\vert j\right\rangle $. \ Of course,
finding a disagreement is what causes him to reject.

The first step, though, is to lower-bound the probability that Arthur finds a
collision. \ Let $\left\vert \varphi\right\rangle =\alpha_{1}\left\vert
1\right\rangle +\cdots+\alpha_{N}\left\vert N\right\rangle $. \ Then for every
copy of $\left\vert \varphi\right\rangle $ and every edge $\left(  i,j\right)
\in\mathcal{M}$, Arthur measures an outcome of the form $\left\vert
i\right\rangle \pm\left\vert j\right\rangle $\ with probability $\left\vert
\alpha_{i}\right\vert ^{2}+\left\vert \alpha_{j}\right\vert ^{2}$, and these
outcomes are independent from one copy to the next. \ We are interested in the
probability that, for some $\left(  i,j\right)  $\ pair, Arthur measures
$\left\vert i\right\rangle \pm\left\vert j\right\rangle $\ more than once.
\ But this is just the famous Birthday Paradox, with $K=\Theta(\sqrt{N}%
)$\ \textquotedblleft people\textquotedblright\ (the copies of $\left\vert
\varphi\right\rangle $) and $N/2$\ \textquotedblleft days\textquotedblright%
\ (the edges in $\mathcal{M}$). \ The one twist is that the distribution over
birthdays need not be uniform. \ However, a result of Bloom and Knight
\cite{bloom} shows that the\ Birthday Paradox occurs in the nonuniform case as well:

\begin{lemma}
[Generalized Birthday Paradox \cite{bloom}]\label{birthday}Suppose there are
$N$ days in the year,\ and each person's birthday is drawn independently from
the same distribution (not necessarily uniform). \ Then if there are
$\Theta(\sqrt{N})$\ people, at least two of them share a birthday
with\ $\Omega\left(  1\right)  $\ probability. \ (Indeed, the probability of a
collision is minimized precisely when the distribution over birthdays is uniform.)
\end{lemma}

Therefore Arthur finds a collision with constant probability. \ The hard part
is to show that he finds a \textit{disagreement} with constant probability.
\ Here, of course, we will have to use the fact that $\left\vert
\varphi\right\rangle $\ is $\varepsilon$-far from proper.

For now, let us restrict attention to two copies of $\left\vert \varphi
\right\rangle $. \ For each edge $\left(  i,j\right)  \in\mathcal{M}$, define
the \textquotedblleft disagreement probability\textquotedblright%
\[
p_{ij}=\frac{2\left\vert \frac{\alpha_{i}+\alpha_{j}}{\sqrt{2}}\right\vert
^{2}\left\vert \frac{\alpha_{i}-\alpha_{j}}{\sqrt{2}}\right\vert ^{2}}{\left(
\left\vert \alpha_{i}\right\vert ^{2}+\left\vert \alpha_{j}\right\vert
^{2}\right)  ^{2}}%
\]
to be the probability that, conditioned on measuring two outcomes of the form
$\left\vert i\right\rangle \pm\left\vert j\right\rangle $, one of the outcomes
is $\left\vert i\right\rangle +\left\vert j\right\rangle $\ and the other one
is $\left\vert i\right\rangle -\left\vert j\right\rangle $. \ Also, say an
edge $\left(  i,j\right)  \in\mathcal{M}$\ is $c$-\textit{unbalanced with
respect to }$\left\vert \varphi\right\rangle $\ if $p_{ij}\geq c$, and let
$\mathcal{S}_{c}\subseteq\mathcal{M}$\ be the set of $c$-unbalanced edges.
\ Say that a set of edges $\mathcal{S}\subseteq\mathcal{M}$\ is $d$%
-\textit{large with respect to }$\left\vert \varphi\right\rangle $ if%
\[
\sum_{\left(  i,j\right)  \in\mathcal{S}}\left(  \left\vert \alpha
_{i}\right\vert ^{2}+\left\vert \alpha_{j}\right\vert ^{2}\right)  \geq d.
\]
Then the key fact is the following:

\begin{theorem}
\label{matchingthm}Suppose $\left\vert \varphi\right\rangle $\ is
$\varepsilon$-far in trace distance from any proper state. \ Then
$\mathcal{S}_{c}$\ is $d$-large with respect to $\left\vert \varphi
\right\rangle $\ with probability at least $1/3$\ over the choice of
$\mathcal{M}$, for some constants $c$ and $d$ depending on $\varepsilon$.
\end{theorem}

The proof of Theorem \ref{matchingthm}\ is deferred to the next
section.

Assuming Theorem \ref{matchingthm}, we can complete the proof of Claim
\ref{unifclaim},\ and hence of Theorem \ref{symthm}. \ The idea is this: when
Arthur performs the Uniformity Test, simply discard all measurement outcomes
that are not of the form $\left\vert i\right\rangle \pm\left\vert
j\right\rangle $\ for some $\left(  i,j\right)  \in\mathcal{S}_{c}$.
\ Assuming $\mathcal{S}_{c}$ is $d$-large---which it is with constant
probability by Theorem \ref{matchingthm}---with overwhelming probability that
still leaves $\Theta(\sqrt{N})$ \textquotedblleft good\textquotedblright%
\ measurement outcomes. \ Then by Lemma \ref{birthday}, with constant
probability there will be a collision among these good outcomes. \ And by the
definition of $\mathcal{S}_{c}$, any such collision will also be a
disagreement with constant probability, thereby causing Arthur to reject.

\subsection{Unbalanced Edges in Random Matchings\label{MATCH}}

The goal of this section is to prove Theorem \ref{matchingthm}, which we now
restate in a more careful way.

\begin{numlessthm}
There exist constants $c,d>0$ for which the following holds. \ Let $N$ be even
and sufficiently large. \ Suppose the state $\left\vert \varphi\right\rangle
=\alpha_{1}\left\vert 1\right\rangle +\cdots+\alpha_{N}\left\vert
N\right\rangle $\ is $\varepsilon$-far in trace distance from any proper state
(that is, any state of the form $\frac{1}{\sqrt{N}}\sum_{i=1}^{N}\left(
-1\right)  ^{x_{i}}\left\vert i\right\rangle $ where $x_{1},\ldots,x_{N}%
\in\left\{  0,1\right\}  $). \ Let $\mathcal{M}$\ be a matching on $\left[
N\right]  $\ chosen uniformly at random, and let $\mathcal{S}$\ be the set of
edges $\left(  i,j\right)  \in\mathcal{M}$\ that are \textquotedblleft%
$c\varepsilon^{8}$-unbalanced,\textquotedblright\ meaning that%
\[
\left\vert \alpha_{i}^{2}-\alpha_{j}^{2}\right\vert ^{2}\geq2c\varepsilon
^{8}\left(  \left\vert \alpha_{i}\right\vert ^{2}+\left\vert \alpha
_{j}\right\vert ^{2}\right)  ^{2}.
\]
Then%
\[
\sum_{\left(  i,j\right)  \in\mathcal{S}}\left(  \left\vert \alpha
_{i}\right\vert ^{2}+\left\vert \alpha_{j}\right\vert ^{2}\right)  \geq
d\varepsilon^{4}%
\]
with probability at least $1/3$\ over the choice of $\mathcal{M}$.

\end{numlessthm}

Given a state $\left\vert \varphi\right\rangle =\alpha_{1}\left\vert
1\right\rangle +\cdots+\alpha_{N}\left\vert N\right\rangle $, define the
\textit{nonuniformity} of $\left\vert \varphi\right\rangle $\ to be%
\[
\operatorname*{NU}\left(  \left\vert \varphi\right\rangle \right)  :=\frac
{1}{2}\sum_{i=1}^{N}\left\vert \left\vert \alpha_{i}\right\vert ^{2}-\frac
{1}{N}\right\vert .
\]
Intuitively, $\operatorname*{NU}\left(  \left\vert \varphi\right\rangle
\right)  $\ measures whether the distribution induced by measuring $\left\vert
\varphi\right\rangle $\ in the standard basis is close to uniform or not. \ We
will divide the proof of Theorem \ref{matchingthm}\ into two cases: first that
$\operatorname*{NU}\left(  \left\vert \varphi\right\rangle \right)
>\varepsilon^{4}/100$ (the \textquotedblleft nonuniform case\textquotedblright%
), and second that $\operatorname*{NU}\left(  \left\vert \varphi\right\rangle
\right)  \leq\varepsilon^{4}/100$ (the \textquotedblleft uniform
case\textquotedblright).

\subsubsection{The Nonuniform Case\label{NONUNIF}}

We now prove Theorem \ref{matchingthm}\ in the case $\operatorname*{NU}\left(
\left\vert \varphi\right\rangle \right)  >\varepsilon^{4}/100$. \ For
convenience, define $\kappa:=\varepsilon^{4}/100$\ and $p_{i}:=\left\vert
\alpha_{i}\right\vert ^{2}$. \ Then the condition%
\[
\left\vert \alpha_{i}^{2}-\alpha_{j}^{2}\right\vert ^{2}\geq2c\varepsilon
^{8}\left(  \left\vert \alpha_{i}\right\vert ^{2}+\left\vert \alpha
_{j}\right\vert ^{2}\right)  ^{2}%
\]
is equivalent to%
\[
p_{i}^{2}+p_{j}^{2}-2\operatorname{Re}\alpha_{i}^{2}\overline{\alpha_{j}}%
^{2}\geq2c\varepsilon^{8}\left(  p_{i}+p_{j}\right)  ^{2},
\]
which will certainly be true whenever $\left(  p_{i}-p_{j}\right)  ^{2}%
\geq2c\varepsilon^{8}\left(  p_{i}+p_{j}\right)  ^{2}$, or equivalently%
\[
\frac{p_{i}}{p_{j}}+\frac{p_{j}}{p_{i}}\geq\frac{2+2c\varepsilon^{8}%
}{1-2c\varepsilon^{8}}.
\]
(If $p_{i}=0$\ or $p_{j}=0$ then we stipulate that the above inequality
holds.) \ Thus, it suffices to prove the following classical lemma.

\begin{lemma}
\label{andylem}Let $N$\ be even and sufficiently large. \ Let $\left(
p_{1},\ldots,p_{N}\right)  $\ be a probability distribution, and suppose%
\[
\frac{1}{2}\sum_{i=1}^{N}\left\vert p_{i}-\frac{1}{N}\right\vert \geq\kappa.
\]
Let $\mathcal{M}$\ be a uniform random matching on $\left[  N\right]  $, and
let $\mathcal{S}$\ be the set of edges $\left(  i,j\right)  \in\mathcal{M}%
$\ such that $p_{i}/p_{j}+p_{j}/p_{i}\geq2+\kappa^{2}/16$. \ Then%
\[
\sum_{\left(  i,j\right)  \in\mathcal{S}}\left(  p_{i}+p_{j}\right)  \geq
\frac{\kappa}{12}%
\]
with probability at least $1/3$\ over $\mathcal{M}$.
\end{lemma}

Let $H$\ (the \textquotedblleft heavy elements\textquotedblright) be the set
of $i\in\left[  N\right]  $\ such that $p_{i}\geq1/N$, and let $H^{\ast
}\subseteq H$\ (the \textquotedblleft very heavy elements\textquotedblright)
be the set of $i\in\left[  N\right]  $\ such that $p_{i}\geq\frac{1+\kappa
/2}{N}$. \ Let $L$\ (the \textquotedblleft light elements\textquotedblright)
be the set of $i\in\left[  N\right]  $\ such that $p_{i}<1/N$, and let
$L^{\ast}\subseteq L$\ (the \textquotedblleft very light
elements\textquotedblright) be the set of $i\in\left[  N\right]  $\ such that
$p_{i}\leq\frac{1-\kappa/4}{N}$. \ Clearly%
\[
\sum_{i\in H}\left(  p_{i}-\frac{1}{N}\right)  =\sum_{i\in L}\left(  \frac
{1}{N}-p_{i}\right)  =\kappa.
\]
Using this, we can prove two simple facts: that there are $\Omega\left(
N\right)  $\ very light elements, and that the very heavy elements have total
weight $\Omega\left(  \kappa\right)  $.

\begin{proposition}
\label{light}$\left\vert L^{\ast}\right\vert \geq\kappa N/2$.
\end{proposition}

\begin{proof}
We have%
\[
\kappa=\sum_{i\in L}\left(  \frac{1}{N}-p_{i}\right)  \leq\frac{\left\vert
L^{\ast}\right\vert }{N}+\left(  \left\vert L\right\vert -\left\vert L^{\ast
}\right\vert \right)  \frac{\kappa}{4N}.
\]
Now use $\left\vert L\right\vert \leq N$\ and rearrange.
\end{proof}

Given any subset $A\subseteq\left[  N\right]  $, define the \textquotedblleft
weight\textquotedblright\ of $A$ to be $W_{A}:=\sum_{i\in A}p_{i}$.

\begin{proposition}
\label{heavy}$W_{H^{\ast}}\geq\kappa/2$.
\end{proposition}

\begin{proof}
We have%
\begin{align*}
\kappa &  =\sum_{i\in H}\left(  p_{i}-\frac{1}{N}\right) \\
&  =\sum_{i\in H\setminus H^{\ast}}\left(  p_{i}-\frac{1}{N}\right)
+\sum_{i\in H^{\ast}}\left(  p_{i}-\frac{1}{N}\right) \\
&  \leq N\frac{\kappa}{2N}+W_{H^{\ast}}.
\end{align*}
Now subtract $\kappa/2$ from both sides.
\end{proof}

To prove Lemma \ref{andylem}, we divide into two cases.

The first case is that $\left\vert H\right\vert \geq N/2$\ (in other words, at
least half the elements are heavy). \ In this case, we begin constructing the
matching $\mathcal{M}$\ by randomly assigning partners to the
\textquotedblleft very light elements\textquotedblright\ $i\in L^{\ast}$.
\ Recall from Proposition \ref{light} that $\left\vert L^{\ast}\right\vert
\geq\kappa N/2$. \ So by a standard Chernoff bound, it is easy to see that at
least (say) $\left\vert L^{\ast}\right\vert /6$\ elements $i\in L^{\ast}%
$\ will be matched to partners $j\in H$, with probability $1-o\left(
1\right)  $ over $\mathcal{M}$. \ Notice that every edge $\left(  i,j\right)
$\ with $i\in L^{\ast}$\ and $j\in H$\ satisfies $p_{i}\leq\frac{1-\kappa
/4}{N}$\ and $p_{j}\geq1/N$, and therefore%
\[
\frac{p_{i}}{p_{j}}+\frac{p_{j}}{p_{i}}\geq1-\kappa/4+\frac{1}{1-\kappa
/4}>2+\frac{\kappa^{2}}{16}.
\]
Thus, all of these edges go into the set $\mathcal{S}$. \ We then have%
\[
\sum_{\left(  i,j\right)  \in\mathcal{S}}\left(  p_{i}+p_{j}\right)  \geq
\frac{\left\vert L^{\ast}\right\vert }{6}\cdot\frac{1}{N}\geq\frac{\kappa}{12}%
\]
and are done.

The second case is that $\left\vert H\right\vert <N/2$\ (in other words, there
are more light elements than heavy ones). \ In this case, we begin
constructing $\mathcal{M}$\ by randomly assigning partners to the
\textquotedblleft very heavy elements\textquotedblright\ $i\in H^{\ast}$.
\ Let $B$\ be the set of elements\ $i\in H^{\ast}$\ that get matched to
partners in $H$. \ Then since $\left\vert H\right\vert <N/2$, every element of
$H^{\ast}$\ goes into $B$ with probability less than $1/2$, and hence%
\[
\operatorname*{E}_{\mathcal{M}}\left[  W_{B}\right]  <\sum_{i\in H^{\ast}%
}\frac{p_{i}}{2}=\frac{W_{H^{\ast}}}{2}.
\]
Therefore%
\[
\Pr_{\mathcal{M}}\left[  W_{B}>\frac{3}{4}W_{H^{\ast}}\right]  <\frac{2}{3}%
\]
by Markov's inequality. \ In other words, with probability greater than $1/3$,
at least $1/4$\ of the probability weight in $H^{\ast}$\ gets matched to
partners in $L$. \ Suppose this happens.

Notice that every edge $\left(  i,j\right)  $ with $i\in H^{\ast}$\ and $j\in
L$\ satisfies $p_{i}\geq\frac{1+\kappa/2}{N}$ and $p_{j}<1/N$, and therefore%
\[
\frac{p_{i}}{p_{j}}+\frac{p_{j}}{p_{i}}>1+\kappa/2+\frac{1}{1+\kappa
/2}>2+\frac{\kappa^{2}}{8}.
\]
Thus, all of these edges go into the set $\mathcal{S}$. \ Furthermore, by the
assumption $W_{B}\leq\frac{3}{4}W_{H^{\ast}}$, we have%
\[
\sum_{\left(  i,j\right)  \in\mathcal{S}}\left(  p_{i}+p_{j}\right)  \geq
\sum_{i\in H^{\ast}\setminus B}p_{i}=W_{H^{\ast}\setminus B}\geq
\frac{W_{H^{\ast}}}{4}\geq\frac{\kappa}{8}%
\]
and are done.

\subsubsection{The Uniform Case\label{UNIF}}

We now prove Theorem \ref{matchingthm}\ for states $\left\vert \varphi
\right\rangle =\alpha_{1}\left\vert 1\right\rangle +\cdots+\alpha
_{N}\left\vert N\right\rangle $\ such that $\operatorname*{NU}\left(
\left\vert \varphi\right\rangle \right)  \leq\varepsilon^{4}/100$. \ The first
step is to define a measure of the distance from $\left\vert \varphi
\right\rangle $\ to the closest proper state, which we call the
\textit{impropriety} of $\left\vert \varphi\right\rangle $\ or
$\operatorname*{imp}\left(  \left\vert \varphi\right\rangle \right)  $:%
\[
\operatorname*{imp}\left(  \left\vert \varphi\right\rangle \right)
:=\min_{\left\vert r\right\vert =1/N}\sum_{i=1}^{N}\left\vert \alpha_{i}%
^{2}-r\right\vert .
\]
Clearly $0\leq\operatorname*{imp}\left(  \left\vert \varphi\right\rangle
\right)  \leq2$ for all $\left\vert \varphi\right\rangle $, with
$\operatorname*{imp}\left(  \left\vert \varphi\right\rangle \right)  =0$ if
and only if $\left\vert \varphi\right\rangle $ is equivalent to a proper state
up to a phase shift. \ We also have the following:

\begin{lemma}
\label{implem}Suppose $\left\vert \varphi\right\rangle $\ is $\varepsilon$-far
in trace distance from any proper state. \ Then $\operatorname*{imp}\left(
\left\vert \varphi\right\rangle \right)  >\varepsilon^{2}$.
\end{lemma}

\begin{proof}
By Proposition \ref{fidelityprop}, we have $\left\vert \left\langle
\varphi|\psi\right\rangle \right\vert
<\sqrt{1-\varepsilon^{2}}<1-\varepsilon ^{2}/2$ for all proper
states $\left\vert \psi\right\rangle $. \ On the other hand, suppose
$\operatorname*{imp}\left(  \left\vert \varphi\right\rangle \right)
\leq\varepsilon^{2}$. \ Then we will construct a proper state
$\left\vert \psi\right\rangle $\ such that $\left\vert \left\langle
\varphi|\psi\right\rangle \right\vert \geq1-\varepsilon^{2}/2$,
thereby obtaining the desired contradiction.

Let $r$ be a complex number with $\left\vert r\right\vert =1/N$\
that minimizes $\sum_{i=1}^{N}\left\vert \alpha_{i}^{2}-r\right\vert
$, and let $\sqrt{r}$\ be a canonical square root of $r$. \ Also let
$\beta_{i}:=\left\vert \alpha_{i}^{2}-r\right\vert $.
\ Then%
\[
\left\vert \alpha_{i}+\sqrt{r}\right\vert \left\vert \alpha_{i}-\sqrt
{r}\right\vert =\left\vert \alpha_{i}^{2}-r\right\vert =\beta_{i},
\]
which means that either $\left\vert \alpha_{i}+\sqrt{r}\right\vert \leq
\sqrt{\beta_{i}}$ or $\left\vert \alpha_{i}-\sqrt{r}\right\vert \leq
\sqrt{\beta_{i}}$. \ So by setting the $\gamma_{i}$'s to $\sqrt{r}$\ or
$-\sqrt{r}$\ appropriately, we can construct a state $\left\vert
\psi\right\rangle =\gamma_{1}\left\vert 1\right\rangle +\cdots+\gamma
_{N}\left\vert N\right\rangle $ that is proper up to a trivial phase factor,
such that $\left\vert \alpha_{i}-\gamma_{i}\right\vert \leq\sqrt{\beta_{i}}%
$\ for all $i$. \ Then%
\begin{align*}
2-2\left\vert \left\langle \varphi|\psi\right\rangle \right\vert  &
=2-2\left\vert \sum_{i=1}^{N}\alpha_{i}\overline{\gamma_{i}}\right\vert \\
&  \leq2-\left\vert \sum_{i=1}^{N}\alpha_{i}\overline{\gamma_{i}}+\sum
_{i=1}^{N}\overline{\alpha_{i}}\gamma_{i}\right\vert \\
&  \leq\left\vert 2-\sum_{i=1}^{N}\alpha_{i}\overline{\gamma_{i}}-\sum
_{i=1}^{N}\overline{\alpha_{i}}\gamma_{i}\right\vert \\
&  =\left\vert \sum_{i=1}^{N}\left(  \left\vert \alpha_{i}\right\vert
^{2}+\left\vert \gamma_{i}\right\vert ^{2}-\alpha_{i}\overline{\gamma_{i}%
}-\overline{\alpha_{i}}\gamma_{i}\right)  \right\vert \\
&  =\sum_{i=1}^{N}\left\vert \alpha_{i}-\gamma_{i}\right\vert ^{2}\\
&  \leq\sum_{i=1}^{N}\beta_{i}\\
&  =\operatorname*{imp}\left(  \left\vert \varphi\right\rangle \right) \\
&  \leq\varepsilon^{2},
\end{align*}
and hence $\left\vert \left\langle \varphi|\psi\right\rangle \right\vert
\geq1-\varepsilon^{2}/2$\ as claimed.
\end{proof}

In what follows, assume $\operatorname*{imp}\left(  \left\vert \varphi
\right\rangle \right)  >\varepsilon^{2}$.

Now as in Section \ref{NONUNIF}, let $p_{i}:=\left\vert \alpha_{i}\right\vert
^{2}$, and for any subset $A\subseteq\left[  N\right]  $, define the
\textquotedblleft probability weight\textquotedblright\ of $A$\ to be
$W_{A}:=\sum_{i\in A}p_{i}$. \ Also, let $\delta:=\varepsilon^{2}/5$, and let
$U$\ (the \textquotedblleft$\delta$-uniform subset\textquotedblright) be the
set of all $i\in\left[  N\right]  $\ such that $\left\vert p_{i}%
-1/N\right\vert \leq\delta/N$. \ The following proposition shows that $U$
encompasses \textquotedblleft most\textquotedblright\ of $\left\vert
\varphi\right\rangle $, whether in terms of cardinality or in terms of
probability weight.

\begin{proposition}
\label{bprop}$\left\vert U\right\vert \geq N\left(  1-\varepsilon
^{2}/10\right)  $ and $W_{U}\geq1-3\varepsilon^{2}/10$.
\end{proposition}

\begin{proof}
We have%
\[
\frac{\varepsilon^{4}}{100}\geq\operatorname*{NU}\left(  \left\vert
\varphi\right\rangle \right)  \geq\frac{1}{2}\left(  N-\left\vert U\right\vert
\right)  \frac{\delta}{N},
\]
hence%
\[
\left\vert U\right\vert \geq N\left(  1-\frac{\varepsilon^{4}}{50\delta
}\right)  =N\left(  1-\frac{\varepsilon^{2}}{10}\right)  ,
\]
hence%
\[
W_{U}\geq N\left(  1-\frac{\varepsilon^{2}}{10}\right)  \left(  \frac{1}%
{N}-\frac{\delta}{N}\right)  \geq1-\frac{3\varepsilon^{2}}{10}.
\]

\end{proof}

Let%
\[
\operatorname*{imp}\nolimits_{U}\left(  \left\vert \varphi\right\rangle
\right)  :=\min_{\left\vert r\right\vert =1/N}\sum_{i\in U}\left\vert
\alpha_{i}^{2}-r\right\vert
\]
be an analogue of impropriety that is restricted to the set $U$. \ By
combining Lemma \ref{implem} with Proposition \ref{bprop}, we can now
lower-bound $\operatorname*{imp}\nolimits_{U}\left(  \left\vert \varphi
\right\rangle \right)  $.

\begin{proposition}
\label{impbprop}$\operatorname*{imp}\nolimits_{U}\left(  \left\vert
\varphi\right\rangle \right)  \geq3\varepsilon^{2}/5.$
\end{proposition}

\begin{proof}
For all $r$ with $\left\vert r\right\vert =1/N$, we have%
\[
\sum_{i\notin U}\left\vert \alpha_{i}^{2}-r\right\vert \leq\sum_{i\notin
U}p_{i}+\sum_{i\notin U}\frac{1}{N}\leq\left(  1-W_{U}\right)  +\frac
{N-\left\vert U\right\vert }{N}\leq\frac{3\varepsilon^{2}}{10}+\frac
{\varepsilon^{2}}{10}=\frac{2\varepsilon^{2}}{5}%
\]
by Proposition \ref{bprop}. \ Hence%
\begin{align*}
\operatorname*{imp}\nolimits_{U}\left(  \left\vert \varphi\right\rangle
\right)   &  =\min_{\left\vert r\right\vert =1/N}\sum_{i\in U}\left\vert
\alpha_{i}^{2}-r\right\vert \\
&  \geq\min_{\left\vert r\right\vert =1/N}\sum_{i\in\left[  N\right]
}\left\vert \alpha_{i}^{2}-r\right\vert -\max_{\left\vert r\right\vert
=1/N}\sum_{i\notin U}\left\vert \alpha_{i}^{2}-r\right\vert \\
&  \geq\operatorname*{imp}\left(  \left\vert \varphi\right\rangle \right)
-\frac{2\varepsilon^{2}}{5}\\
&  \geq\frac{3\varepsilon^{2}}{5}.
\end{align*}

\end{proof}

We are finally ready for the geometric core of our result. \ Let $V$\ be a
collection of vectors in $\mathbb{R}^{2}$ (possibly with multiplicity), which
consists of the vector $\left(  N\operatorname{Re}\alpha_{i}^{2}%
,N\operatorname{Im}\alpha_{i}^{2}\right)  $\ for every $i\in U$. \ Let
$\left\Vert v\right\Vert $\ be the $2$-norm of $v$. \ Then we know by the
definition of $U$ that $1-\delta\leq\left\Vert v\right\Vert \leq1+\delta$ for
all $v\in V$. \ We also know from Proposition \ref{bprop}\ that%
\[
\left\vert V\right\vert =\left\vert U\right\vert \geq N\left(  1-\frac
{\varepsilon^{2}}{10}\right)  \geq0.9N,
\]
and from Proposition \ref{impbprop} that for all unit vectors $w\in
\mathbb{R}^{2}$,%
\[
\sum_{v\in V}\left\Vert v-w\right\Vert \geq\frac{3\varepsilon^{2}N}{5}%
\geq\frac{3\varepsilon^{2}\left\vert V\right\vert }{5}.
\]
Based on this information, we want to find two subsets $X,Y\subseteq V$, both
of size $\Omega\left(  \left\vert V\right\vert \right)  $, such that
$\left\Vert x-y\right\Vert =\Omega\left(  1\right)  $ for all $x\in X$\ and
$y\in Y$.

For suppose we can do this. \ Then just as in Section \ref{NONUNIF}, when a
matching $\mathcal{M}$\ on $\left[  N\right]  $ is chosen uniformly at random,
by a Chernoff bound it will have $\Omega\left(  N\right)  $\ edges between the
subsets of $\left[  N\right]  $\ corresponding to $X$\ and $Y$\ with
overwhelming probability. \ Assuming that happens,\ we will have $\left\vert
\alpha_{i}^{2}-\alpha_{j}^{2}\right\vert =\Omega\left(  1/N\right)  $ for
every such edge $\left(  i,j\right)  \in\mathcal{M}$, and hence all of these
edges will get added to the set $\mathcal{S}$. \ We will therefore have%
\[
\sum_{\left(  i,j\right)  \in\mathcal{S}}\left(  p_{i}+p_{j}\right)
=\Omega\left(  1\right)
\]
as desired. \ (For simplicity, we have suppressed the dependence on
$\varepsilon$ here.)

What we need, then, is the following geometric lemma.

\begin{lemma}
\label{geolem}Let $V$\ be a collection of vectors in the plane. \ Suppose that
$1-\delta\leq\left\Vert v\right\Vert \leq1+\delta$\ for every $v\in V$,\ and
that $\sum_{v\in V}\left\Vert v-w\right\Vert \geq\kappa\left\vert V\right\vert
$ for every unit vector $w\in\mathbb{R}^{2}$. \ Then provided $\delta
\leq\kappa/2$, there exist subsets $X,Y\subseteq V$, both of size at least
$\kappa\left\vert V\right\vert /40$, such that $\left\Vert x-y\right\Vert
\geq\kappa/20$\ for all $x\in X$ and $y\in Y$.\footnote{We did not try to
optimize the constants.}
\end{lemma}

\begin{proof}
Divide the plane into $K\in\left[  30/\kappa,40/\kappa\right]  $\ equal-sized,
half-open angular sectors, centered about the origin. \ By the pigeonhole
principle, one of these sectors (call it $S$) must contain at least
$\left\vert V\right\vert /K\geq\kappa\left\vert V\right\vert /40$\ of the
vectors. Let $S^{\prime}$ be the union of $S$ and its two adjacent sectors.
\ Then we claim that at least $\kappa\left\vert V\right\vert /40$\ of the
vectors must lie outside of $S^{\prime}$. \ For suppose not. \ Then let
$z$\ be the unit vector that bisects $S$, and let%
\[
\theta=\frac{3}{2}\left(  \frac{2\pi}{K}\right)  \leq\frac{3}{2}\left(
\frac{2\pi}{30/\kappa}\right)  =\frac{\pi\kappa}{10}%
\]
be the angle between $z$\ and the border of $S^{\prime}$. \ Notice that by the
triangle inequality, we have%
\[
\left\Vert v-z\right\Vert \leq\sqrt{2-2\cos\theta}+\delta\leq\theta+\delta
\leq\frac{\pi\kappa}{10}+\delta
\]
for every $v$ in $S^{\prime}$ (where we have used the bound $\cos\theta
\geq1-\theta^{2}/2$). \ We also have $\left\Vert v-z\right\Vert \leq2+\delta$
for every $v\in V$. \ Hence%
\begin{align*}
\sum_{v\in V}\left\Vert v-z\right\Vert  &  \leq\sum_{v\in S^{\prime}}\left(
\frac{\pi\kappa}{10}+\delta\right)  +\sum_{v\notin S^{\prime}}\left(
2+\delta\right) \\
&  \leq\frac{\pi\kappa}{10}\left\vert V\right\vert +2\frac{\kappa\left\vert
V\right\vert }{40}+\delta\left\vert V\right\vert \\
&  <\kappa\left\vert V\right\vert
\end{align*}
which is a contradiction.

Now let $X$ be the set of all $v$'s in $S$, and let $Y$\ be the set
of all $v$'s outside $S^{\prime}$. \ Then $\left\vert X\right\vert
\geq\kappa\left\vert V\right\vert /40$\ and $\left\vert
Y\right\vert \geq\kappa\left\vert V\right\vert /40$. \ Also, let%
\[
\tau=\frac{2\pi}{K}\geq\frac{\pi\kappa}{20}%
\]
be the angle of a single sector. \ Then it is not hard to see that for all
$x\in X$ and $y\in Y$,%
\[
\left\Vert x-y\right\Vert \geq\left(  1-\delta\right)  \sqrt{2-2\cos\tau}%
\geq\left(  1-\delta\right)  \frac{\tau}{\sqrt{2}}\geq\left(  1-\frac{\kappa
}{2}\right)  \frac{\pi\kappa}{20\sqrt{2}}\geq\frac{\kappa}{20}%
\]
where we have used the bound $\cos\tau\leq1-\tau^{2}/4$\ for all $\tau
\in\left[  0,\pi/2\right]  $.
\end{proof}

Now set $\kappa:=3\varepsilon^{2}/5$. \ Then $\delta=\varepsilon^{2}%
/5<\kappa/2$ and the condition of Lemma \ref{geolem} is satisfied. \ So
considering the sets $X,Y$\ from the lemma, we have%
\[
\left\vert X\right\vert ,\left\vert Y\right\vert \geq\frac{\kappa\left\vert
V\right\vert }{40}\geq\frac{0.9\kappa N}{40}=\frac{\left(  0.9\right)
3\varepsilon^{2}N}{200}>\frac{\varepsilon^{2}N}{100},
\]
and also%
\[
\left\Vert x-y\right\Vert \geq\frac{\kappa}{20}\geq\frac{3\varepsilon^{2}%
}{100}%
\]
for all $x\in X$ and $y\in Y$. \ This means that we can find subsets
$X^{\prime},Y^{\prime}\subseteq\left[  N\right]  $\ such that

\begin{enumerate}
\item[(i)] $\left\vert X^{\prime}\right\vert ,\left\vert Y^{\prime}\right\vert
\geq\varepsilon^{2}N/100$\ and

\item[(ii)] $\left\vert \alpha_{i}^{2}-\alpha_{j}^{2}\right\vert \geq
\frac{3\varepsilon^{2}}{100N}$ for all $i\in X^{\prime}$\ and $j\in Y^{\prime
}$.
\end{enumerate}

Property (ii) implies that%
\[
\left\vert \alpha_{i}^{2}-\alpha_{j}^{2}\right\vert ^{2}\geq\frac
{9\varepsilon^{4}}{10000N^{2}}\geq2c\varepsilon^{8}\left(  \left\vert
\alpha_{i}\right\vert ^{2}+\left\vert \alpha_{j}\right\vert ^{2}\right)  ^{2}%
\]
for some suitable constant $c$. \ Hence every edge $\left(  i,j\right)
\in\mathcal{M}$\ with $i\in X^{\prime}$\ and $j\in Y^{\prime}$\ will get added
to the set $\mathcal{S}$ (again assuming a suitable $c$).

Property (i), together with a Chernoff bound, implies that with probability
$1-o\left(  1\right)  $\ over the choice of matching $\mathcal{M}$, there are
at least (say) $\varepsilon^{4}N/20000$\ edges $\left(  i,j\right)
\in\mathcal{M}$\ such that $i\in X^{\prime}$\ and $j\in Y^{\prime}$. \ Suppose
this happens. \ Then%
\[
\sum_{\left(  i,j\right)  \in\mathcal{S}}\left(  p_{i}+p_{j}\right)  \geq
\frac{\varepsilon^{4}N}{20000}\cdot2\left(  \frac{1-\delta}{N}\right)
=\Omega\left(  \varepsilon^{4}\right)
\]
as desired. \ This completes the proof of Theorem \ref{matchingthm}.

\subsection{The General Case\label{GENERAL}}

Of course, in general the states $\left\vert \varphi_{1}\right\rangle
,\ldots,\left\vert \varphi_{K}\right\rangle $ sent by the $K=\Theta(\sqrt{N})$
Merlins need not be identical. \ To deal with this, we now give our final
protocol, which removes the symmetry restriction.

\begin{center}
\fbox{ \begin{minipage}{5.5in}
\textbf{The }\textsc{2-out-of-4-Sat}\textbf{ Protocol} \medskip

Given $\left\vert \varphi_{1}\right\rangle ,\ldots,\left\vert
\varphi _{K}\right\rangle $, Arthur performs one of the following
three tests, each with probability $1/3$. \medskip

\textbf{Satisfiability Test:} Arthur applies the Satisfiability
Test, described in Section \ref{PROPER}, to $\left\vert
\varphi_{1}\right\rangle $.
\medskip

\textbf{Symmetry Test:} Arthur chooses an index $k\in\left\{
2,\ldots ,K\right\}  $ uniformly at random, performs a swap test
between $\left\vert \varphi_{1}\right\rangle $\ and $\left\vert
\varphi_{k}\right\rangle $, and accepts if and only if the swap test
accepts. \medskip

\textbf{Uniformity Test:} Arthur chooses a matching $\mathcal{M}$\
on $\left[ N\right]  $ uniformly at random. \ He then measures each\
$\left\vert \varphi_{k}\right\rangle $\ in an orthonormal basis,
which contains the vectors\[ \frac{\left\vert i\right\rangle
+\left\vert j\right\rangle }{\sqrt{2}},\frac{\left\vert
i\right\rangle -\left\vert j\right\rangle }{\sqrt{2}}\] for every
edge $\left(  i,j\right) \in\mathcal{M}$. \ If for some $\left(
i,j\right)  \in\mathcal{M}$, the two outcomes $\frac{\left\vert
i\right\rangle +\left\vert j\right\rangle }{\sqrt{2}}$\ and
$\frac{\left\vert i\right\rangle -\left\vert j\right\rangle
}{\sqrt{2}}$\ both occur among the $K$ measurement outcomes, then
Arthur rejects. \ Otherwise he accepts.
\end{minipage}}
\end{center}

It is clear that the above protocol has perfect completeness, and thus the
problem is to show soundness: that is, if $\phi$\ is unsatisfiable, then one
of the three tests rejects with constant probability. \ There are three cases.

The first case is that $\left\vert \varphi_{1}\right\rangle $ is $\varepsilon
$-close to some proper state $\left\vert \psi\right\rangle $. \ Then as
before, the Satisfiability Test will reject with constant probability,
provided we choose $\varepsilon$ sufficiently small.

The second case is that $\left\vert \left\langle \varphi_{1}|\varphi
_{k}\right\rangle \right\vert <1-\delta$\ for at least a $\gamma$\ fraction of
indices $k\in\left\{  2,\ldots,K\right\}  $. \ In that case it is clear that
the Symmetry Test will reject with probability at least $\gamma\delta/2$.

The third case is that $\left\vert \left\langle \varphi_{1}|\varphi
_{k}\right\rangle \right\vert \geq1-\delta$\ for more than a $1-\gamma
$\ fraction of indices $k\in\left\{  2,\ldots,K\right\}  $,\ but nevertheless
$\left\vert \varphi_{1}\right\rangle $\ is $\varepsilon$-far from any proper
state. \ In this case we need to generalize the results of the previous
section, to show that the Uniformity Test will still reject with constant
probability (dependent on $\varepsilon$, $\delta$, and $\gamma$).

The first step in the analysis is simply to discard all states $\left\vert
\varphi_{k}\right\rangle $\ such that $\left\vert \left\langle \varphi
_{1}|\varphi_{k}\right\rangle \right\vert <1-\delta$. \ By Proposition
\ref{fidelityprop}, the remaining $K^{\prime}\geq\left(  1-\gamma\right)
K$\ states are all $\sqrt{2\delta}$-close to $\left\vert \varphi
_{1}\right\rangle $\ in trace distance.

Now given a matching $\mathcal{M}$ on $\left[  N\right]  $,\ let
$\mathcal{S}_{c}$\ be the set of edges in $\mathcal{M}$\ that are
$c$-unbalanced with respect to $\left\vert \varphi_{1}\right\rangle $, in the
sense defined in Section \ref{SYMMETRIC}. \ Then Theorem \ref{matchingthm}
implies that $\mathcal{S}_{c}$\ is $d$-large with respect to $\left\vert
\varphi_{1}\right\rangle $ (for some constants $c$ and $d$) with probability
at least $1/3$\ over the choice of $\mathcal{M}$. \ Suppose that it is.

Call a measurement outcome $\left\vert i\right\rangle \pm\left\vert
j\right\rangle $ \textit{good} if $\left(  i,j\right)  \in\mathcal{S}_{c}$.
\ Then when Arthur performs the Uniformity Test, we simply discard all states
for which the outcome is not good. \ Since all of the states are
$\sqrt{2\delta}$-close to $\left\vert \varphi_{1}\right\rangle $, and since
$\mathcal{S}_{c}$\ is $d$-large with respect to $\left\vert \varphi
_{1}\right\rangle $, with overwhelming probability this still leaves us with
$K^{\prime\prime}\approx(d-\sqrt{2\delta})K^{\prime}$\ states. \ Call those
states $\left\vert \xi_{1}\right\rangle ,\ldots,\left\vert \xi_{K^{\prime
\prime}}\right\rangle $.

Let $\widetilde{\mathcal{M}}=\left\{  \left\vert i\right\rangle \pm\left\vert
j\right\rangle :\left(  i,j\right)  \in\mathcal{M}\right\}  $. \ Given a state
$\left\vert \varphi\right\rangle $, let $\mathcal{D}_{\left\vert
\varphi\right\rangle }$\ be the probability distribution over $\widetilde
{\mathcal{M}}$\ induced by measuring $\left\vert \varphi\right\rangle
$\ according to $\mathcal{M}$. \ Then we know that $\left\Vert \mathcal{D}%
_{\left\vert \varphi_{1}\right\rangle }-\mathcal{D}_{\left\vert \xi
_{k}\right\rangle }\right\Vert \leq\sqrt{2\delta}$ for all $k\in\left[
K^{\prime\prime}\right]  $. \ Next let $\mathcal{D}_{\left\vert \varphi
\right\rangle }^{\prime}$\ be the distribution over $\widetilde{\mathcal{M}}$
induced by measuring $\left\vert \varphi\right\rangle $, and then conditioning
on the outcome being good. \ Then we claim that%
\[
\left\Vert \mathcal{D}_{\left\vert \xi_{k}\right\rangle }^{\prime}%
-\mathcal{D}_{\left\vert \varphi_{1}\right\rangle }^{\prime}\right\Vert
\leq\frac{\sqrt{2\delta}}{d-\sqrt{2\delta}}%
\]
for all $k\in\left[  K^{\prime\prime}\right]  $, where as always $\left\Vert
\cdot\right\Vert $\ denotes the variation distance. \ This is so because of
the following simple fact:

\begin{proposition}
\label{condvar}Let $\mathcal{D}_{1}$\ and $\mathcal{D}_{2}$\ be probability
distributions, let $E$ be an event, and let $\mathcal{D}_{1}^{\prime}$\ and
$\mathcal{D}_{2}^{\prime}$\ denote $\mathcal{D}_{1}$\ and $\mathcal{D}_{2}%
$\ respectively\ conditioned on $E$. \ Suppose $\left\Vert \mathcal{D}%
_{1}-\mathcal{D}_{2}\right\Vert \leq\kappa$ and $\Pr_{x\in\mathcal{D}_{1}%
}\left[  E\left(  x\right)  \right]  \geq a$. \ Then $\left\Vert
\mathcal{D}_{1}^{\prime}-\mathcal{D}_{2}^{\prime}\right\Vert \leq\frac{\kappa
}{a-\kappa}$.
\end{proposition}

\begin{proof}
Let $b=\Pr_{x\in\mathcal{D}_{2}}\left[  E\left(  x\right)  \right]  $, and
note that $\left\vert a-b\right\vert \leq\kappa$. \ Also let $p_{x}%
=\Pr_{\mathcal{D}_{1}}\left[  x\right]  $ and $q_{x}=\Pr_{\mathcal{D}_{2}%
}\left[  x\right]  $. \ Then%
\begin{align*}
\left\Vert \mathcal{D}_{1}^{\prime}-\mathcal{D}_{2}^{\prime}\right\Vert  &
=\frac{1}{2}\sum_{x:E\left(  x\right)  }\left\vert \Pr_{\mathcal{D}%
_{1}^{\prime}}\left[  x\right]  -\Pr_{\mathcal{D}_{2}^{\prime}}\left[
x\right]  \right\vert \\
&  =\frac{1}{2}\sum_{x:E\left(  x\right)  }\left\vert \frac{p_{x}}{a}%
-\frac{q_{x}}{b}\right\vert \\
&  \leq\frac{1}{2b}\sum_{x:E\left(  x\right)  }\left(  \left\vert p_{x}%
-q_{x}\right\vert +\left\vert p_{x}-\frac{b}{a}p_{x}\right\vert \right) \\
&  \leq\frac{\kappa}{2b}+\frac{1}{2}\left\vert 1-\frac{a}{b}\right\vert \\
&  \leq\frac{\kappa}{b}\\
&  \leq\frac{\kappa}{a-\kappa}.
\end{align*}

\end{proof}

By construction, every measurement outcome $\left\vert i\right\rangle
\pm\left\vert j\right\rangle $\ in the support of every $\mathcal{D}%
_{\left\vert \xi_{k}\right\rangle }^{\prime}$\ corresponds to an edge $\left(
i,j\right)  $ that is $c$-unbalanced with respect to $\left\vert \varphi
_{1}\right\rangle $. \ But this still leaves a key question unanswered: is
$\left(  i,j\right)  $\ reasonably unbalanced with respect to $\left\vert
\xi_{k}\right\rangle $ itself? \ The following lemma will imply that it is,
with high probability over $\mathcal{D}_{\left\vert \xi_{k}\right\rangle
}^{\prime}$: in particular that%
\[
\Pr_{\left\vert i\right\rangle \pm\left\vert j\right\rangle \in\mathcal{D}%
_{\left\vert \xi_{k}\right\rangle }^{\prime}}\left[  \left(  i,j\right)
\text{ is }\frac{c}{4}\text{-unbalanced w.r.t. }\left\vert \xi_{k}%
\right\rangle \right]  \geq1-\frac{16\sqrt{2\delta}}{c\left(  d-\sqrt{2\delta
}\right)  }%
\]
for all $k\in\left[  K^{\prime\prime}\right]  $.

\begin{lemma}
\label{unbalwrt}Let $\mathcal{D}=\left(  p_{x},q_{x}\right)  _{x\in\left[
N\right]  }$\ and $\mathcal{D}^{\prime}=\left(  p_{x}^{\prime},q_{x}^{\prime
}\right)  _{x\in\left[  N\right]  }$\ be any two probability
distributions\ over the set $\left[  N\right]  \times\left\{  0,1\right\}  $.
\ Suppose that $\left\Vert \mathcal{D}-\mathcal{D}^{\prime}\right\Vert \leq
\mu$, and that $2p_{x}q_{x}\geq c\left(  p_{x}+q_{x}\right)  ^{2}$ for every
$x\in\left[  N\right]  $. \ Let $\mathcal{S}$\ be the set of all $x\in\left[
N\right]  $\ such that $2p_{x}^{\prime}q_{x}^{\prime}\geq c^{\prime}\left(
p_{x}^{\prime}+q_{x}^{\prime}\right)  ^{2}$. \ Then%
\[
\sum_{x\in\mathcal{S}}\left(  p_{x}^{\prime}+q_{x}^{\prime}\right)
\geq1-\frac{8\mu}{c-2c^{\prime}},
\]
for all constants $c\in\left(  0,1/2\right)  $\ and $c^{\prime}\in\left(
0,c/2\right)  $.
\end{lemma}

\begin{proof}
Assume for simplicity that $p_{x},q_{x}>0$\ for all $x$ (it is not hard to
remove this restriction). \ Let $\varepsilon_{x}=p_{x}^{\prime}-p_{x}$ and
$\delta_{x}=q_{x}^{\prime}-q_{x}$. \ Then by assumption,%
\[
\sum_{x=1}^{N}\left(  \left\vert \varepsilon_{x}\right\vert +\left\vert
\delta_{x}\right\vert \right)  \leq2\mu.
\]
Let $\overline{\mathcal{S}}$\ be the complement of $\mathcal{S}$. \ Consider
an adversary with a \textquotedblleft budget\textquotedblright\ of $2\mu$, who
is trying to perturb $\mathcal{D}$ so as to maximize $\sum_{x\in
\overline{\mathcal{S}}}\left(  p_{x}^{\prime}+q_{x}^{\prime}\right)  $.
\ Define the \textquotedblleft price per pound\textquotedblright\ of $x$ to be%
\[
\$_{x}:=\frac{\left\vert \varepsilon_{x}\right\vert +\left\vert \delta
_{x}\right\vert }{p_{x}^{\prime}+q_{x}^{\prime}}.
\]
Intuitively, $\$_{x}$\ is the amount the adversary has to \textquotedblleft
spend\textquotedblright\ on perturbing $p_{x}$ and $q_{x}$, divided by the
amount of probability mass that gets added to $\overline{\mathcal{S}}$ as a
result. \ We will show that $\$_{x}\geq\left(  c-2c^{\prime}\right)  /4$\ for
all $x\in\overline{\mathcal{S}}$. \ This will suffice to prove the lemma,
since we then have%
\[
\sum_{x\in\overline{\mathcal{S}}}\left(  p_{x}^{\prime}+q_{x}^{\prime}\right)
=\sum_{x\in\overline{\mathcal{S}}}\frac{\left\vert \varepsilon_{x}\right\vert
+\left\vert \delta_{x}\right\vert }{\$_{x}}\leq\frac{8\mu}{c-2c^{\prime}}.
\]
We now lower-bound $\$_{x}$. \ If we simply divide through by $p_{x}q_{x}$,
the condition $2p_{x}q_{x}\geq c\left(  p_{x}+q_{x}\right)  ^{2}$ is
equivalent to $p_{x}/q_{x}+q_{x}/p_{x}\leq\left(  2-2c\right)  /c$. \ Let
$A=\left(  2-2c\right)  /c$; then in particular, we have $p_{x}\leq Aq_{x}$
and $q_{x}\leq Ap_{x}$\ for all $x$. \ On the other hand, to get
$x\in\overline{\mathcal{S}}$\ we need $p_{x}^{\prime}/q_{x}^{\prime}%
+q_{x}^{\prime}/p_{x}^{\prime}>\left(  2-2c^{\prime}\right)  /c^{\prime}$, and
hence either $p_{x}^{\prime}/q_{x}^{\prime}>B$\ or $q_{x}^{\prime}%
/p_{x}^{\prime}>B$ where $B=\left(  1-c^{\prime}\right)  /c^{\prime}$.

Suppose $p_{x}^{\prime}/q_{x}^{\prime}>B$ without loss of generality. \ Then%
\[
p_{x}+\varepsilon_{x}>B\left(  q_{x}+\delta_{x}\right)  >B\left(  \frac{p_{x}%
}{A}+\delta_{x}\right)  ,
\]
which rearranging means%
\[
\varepsilon_{x}-B\delta_{x}>\left(  \frac{B}{A}-1\right)  p_{x}.
\]
Likewise%
\[
B\left(  q_{x}+\delta_{x}\right)  <p_{x}+\varepsilon_{x}<Aq_{x}+\varepsilon
_{x},
\]
which rearranging means%
\[
\varepsilon_{x}-B\delta_{x}>\left(  B-A\right)  q_{x}>\left(  \frac{B}%
{A}-1\right)  q_{x}.
\]
Combining,%
\[
\varepsilon_{x}-B\delta_{x}>\left(  \frac{B}{A}-1\right)  \frac{p_{x}+q_{x}%
}{2}%
\]
and hence%
\[
\left\vert \varepsilon_{x}\right\vert +\left\vert \delta_{x}\right\vert
>\frac{1}{B}\left(  \varepsilon_{x}-B\delta_{x}\right)  >\left(  \frac{1}%
{A}-\frac{1}{B}\right)  \frac{p_{x}+q_{x}}{2}.
\]
Therefore%
\begin{align*}
\$_{x}  &  =\frac{\left\vert \varepsilon_{x}\right\vert +\left\vert \delta
_{x}\right\vert }{p_{x}^{\prime}+q_{x}^{\prime}}\\
&  \geq\frac{\left\vert \varepsilon_{x}\right\vert +\left\vert \delta
_{x}\right\vert }{p_{x}+q_{x}+\left\vert \varepsilon_{x}\right\vert
+\left\vert \delta_{x}\right\vert }\\
&  >\frac{\frac{1}{2}\left(  1/A-1/B\right)  }{1+\frac{1}{2}\left(
1/A-1/B\right)  }\\
&  =\frac{c/\left(  2-2c\right)  -c^{\prime}/\left(  1-c^{\prime}\right)
}{2+c/\left(  2-2c\right)  -c^{\prime}/\left(  1-c^{\prime}\right)  }\\
&  =\frac{c-2c^{\prime}+cc^{\prime}}{4-3c-6c^{\prime}+5cc^{\prime}}\\
&  \geq\frac{c-2c^{\prime}}{4}%
\end{align*}
as claimed.
\end{proof}

We now need one last conditioning step: discard all states $\left\vert \xi
_{k}\right\rangle $\ for which the measurement outcome is not $c/4$-unbalanced
with respect to $\left\vert \xi_{k}\right\rangle $. \ By Lemma \ref{unbalwrt},
with overwhelming probability this still leaves us with $K^{\prime\prime
\prime}\approx K^{\prime\prime}$\ states (for suitable choices of $c$,\ $d$,
and $\delta$). \ Call those states $\left\vert \varsigma_{1}\right\rangle
,\ldots,\left\vert \varsigma_{K^{\prime\prime\prime}}\right\rangle $.

Given any state $\left\vert \varphi\right\rangle $, let $\mathcal{D}%
_{\left\vert \varphi\right\rangle }^{\prime\prime}$\ be the probability
distribution over $\left\vert i\right\rangle \pm\left\vert j\right\rangle
\in\widetilde{\mathcal{M}}$ obtained by starting from $\mathcal{D}_{\left\vert
\varphi\right\rangle }^{\prime}$, and then conditioning on the edge $\left(
i,j\right)  $\ being $c/4$-unbalanced with respect to $\left\vert
\varphi\right\rangle $. \ Then%
\begin{align*}
\left\Vert \mathcal{D}_{\left\vert \varsigma_{k}\right\rangle }^{\prime\prime
}-\mathcal{D}_{\left\vert \varphi_{1}\right\rangle }^{\prime}\right\Vert  &
\leq\left\Vert \mathcal{D}_{\left\vert \varsigma_{k}\right\rangle }%
^{\prime\prime}-\mathcal{D}_{\left\vert \varsigma_{k}\right\rangle }^{\prime
}\right\Vert +\left\Vert \mathcal{D}_{\left\vert \varsigma_{k}\right\rangle
}^{\prime}-\mathcal{D}_{\left\vert \varphi_{1}\right\rangle }^{\prime
}\right\Vert \\
&  \leq\frac{16\sqrt{2\delta}}{c\left(  d-\sqrt{2\delta}\right)  }+\frac
{\sqrt{2\delta}}{d-\sqrt{2\delta}}\\
&  \leq\frac{17\sqrt{2\delta}}{c\left(  d-\sqrt{2\delta}\right)  }%
\end{align*}
for all $k\in\left[  K^{\prime\prime\prime}\right]  $. \ So by the triangle
inequality,%
\[
\left\Vert \mathcal{D}_{\left\vert \varsigma_{k}\right\rangle }^{\prime\prime
}-\mathcal{D}_{\left\vert \varsigma_{_{\ell}}\right\rangle }^{\prime\prime
}\right\Vert \leq\frac{34\sqrt{2\delta}}{c\left(  d-\sqrt{2\delta}\right)  }%
\]
for all $k,\ell\in\left[  K^{\prime\prime\prime}\right]  $.

So to sum up: we have $K^{\prime\prime\prime}=\Theta(\sqrt{N})$\ samples from
$\widetilde{\mathcal{M}}$, drawn independently from probability distributions
$\mathcal{D}_{\left\vert \varsigma_{1}\right\rangle }^{\prime\prime}%
,\ldots,\mathcal{D}_{\left\vert \varsigma_{K^{\prime\prime\prime}%
}\right\rangle }^{\prime\prime}$\ respectively. \ The distributions
$\mathcal{D}_{\left\vert \varsigma_{k}\right\rangle }^{\prime\prime}$\ have
bounded variation distance from one another. \ We also know, because the
$\mathcal{D}_{\left\vert \varsigma_{k}\right\rangle }^{\prime\prime}$'s only
involve $c/4$-unbalanced edges $\left(  i,j\right)  $, that if Arthur finds a
collision among the $K^{\prime\prime\prime}$\ samples (i.e., two samples of
the form $\left\vert i\right\rangle \pm\left\vert j\right\rangle $\ for some
$\left(  i,j\right)  $), then that collision will also be a disagreement with
constant probability. \ Thus, the one remaining task is to show that Arthur
finds a collision with constant probability.

Showing this amounts to generalizing the Birthday Paradox still further, to
the case where the birthday distributions are not only nonuniform but can also
differ from each other by small amounts. \ In particular we want the following:

\begin{theorem}
\label{birthday2}Let $X_{1},\ldots,X_{K}$\ be independent random variables
over $\left[  N\right]  $, and let $\mathcal{D}_{i}$\ be the distribution over
$X_{i}$. \ Suppose $K\geq32\sqrt{N}$\ and $\left\Vert \mathcal{D}%
_{i}-\mathcal{D}_{j}\right\Vert \leq1/10$ for all $i,j$. \ Then%
\[
\Pr\left[  \exists i,j:X_{i}=X_{j}\right]  \geq\frac{1}{2}.
\]

\end{theorem}

In Section \ref{BDP}, we present a proof of Theorem \ref{birthday2}\ based on
the second moment method. \ (Indeed, our proof works even if the $X_{i}$'s are
only $4$-wise independent.)

By Theorem \ref{birthday2}, Arthur will find a collision among the
$K^{\prime\prime\prime}=\Theta(\sqrt{N})$\ remaining samples with constant
probability. \ Then by the definition of the $\mathcal{D}_{\left\vert
\varsigma_{k}\right\rangle }^{\prime\prime}$'s, this collision will be a
disagreement with constant probability, thereby causing Arthur to reject.

So in summary, we get a protocol with perfect completeness, constant
soundness, and $\widetilde{O}(\sqrt{m})$\ unentangled witnesses with $O\left(
\log m\right)  $\ qubits each.

As a final remark, we can reduce the soundness error to be negligibly small
(in particular, $2^{-\operatorname*{polylog}m}$). \ To do so, we simply
multiply the number of Merlins by a further $\operatorname*{polylog}%
m$\ factor, and repeat the whole protocol $\operatorname*{polylog}m$\ times.

\subsection{The Generalized Birthday Paradox\label{BDP}}

The purpose of this section is to prove the Birthday Paradox, even in the very
general situation where

\begin{enumerate}
\item[(1)] the distributions over birthdays need not be uniform, and

\item[(2)] the distributions need not be the same for every person, but only
$\varepsilon$-close in variation distance, and

\item[(3)] the distributions need not be independent, but only $4$-wise independent.
\end{enumerate}

First we need two lemmas.

\begin{lemma}
\label{epsclose}Let $\mathcal{D}_{1},\mathcal{D}_{2}$\ be probability
distributions over $\left[  n\right]  $ such that $\left\Vert \mathcal{D}%
_{1}-\mathcal{D}_{2}\right\Vert \leq\varepsilon$. \ Then $\Pr_{x\in
\mathcal{D}_{1},y\in\mathcal{D}_{2}}\left[  x=y\right]  \geq\left(
1-\varepsilon\right)  ^{2}/n$.
\end{lemma}

\begin{proof}
Let $p_{x}=\Pr_{\mathcal{D}_{1}}[x]$ and let $q_{x}=\Pr_{\mathcal{D}_{2}}[x]$.
\ Then%
\begin{align*}
\Pr_{x\in\mathcal{D}_{1},y\in\mathcal{D}_{2}}\left[  x=y\right]   &
=\sum_{x\in\left[  n\right]  }p_{x}q_{x}\\
&  \geq\sum_{x\in\left[  n\right]  }\min\left(  p_{x},q_{x}\right)  ^{2}\\
&  \geq\frac{1}{n}\left(  \sum_{x\in\left[  n\right]  }\min\left(  p_{x}%
,q_{x}\right)  \right)  ^{2}\\
&  =\frac{1}{n}\left(  1-\varepsilon\right)  ^{2}%
\end{align*}
where the third line follows from Cauchy-Schwarz.
\end{proof}

\begin{lemma}
\label{sortlem}Let $p_{1},\ldots,p_{K}$ be nonnegative reals, and let
$r=\sum_{i<j<k}p_{i}p_{j}p_{k}$\ and $s=\sum_{i<j}p_{i}p_{j}$. \ Then
$r^{2}\leq2s^{3}$.
\end{lemma}

\begin{proof}
Let $\mathcal{S}$\ be the set of $6$-tuples $\left(  i,j,k,\ell,m,n\right)  $
such that $i<j$, $k<\ell$, and $m<n$, and let $\mathcal{R}$ be the set of
$6$-tuples such that $i<j<k$\ and $\ell<m<n$. \ Then%
\[
s^{3}=\sum_{\mathcal{S}}p_{i}p_{j}p_{k}p_{\ell}p_{m}p_{n}%
\]
while%
\[
r^{2}=\sum_{\mathcal{R}}p_{i}p_{j}p_{k}p_{\ell}p_{m}p_{n}.
\]
Now define a mapping from $\mathcal{R}$\ to $\mathcal{S}$, by simply swapping
$k$\ and $\ell$\ if $k>\ell$, or swapping $\ell$ and $m$ if $k=\ell$. \ It is
easily checked that this mapping is two-to-one. \ Hence $r^{2}\leq2s^{3}$ as claimed.
\end{proof}

We now prove Theorem \ref{birthday2}, which we restate for convenience.

\begin{numlessthm}
Let $X_{1},\ldots,X_{K}$\ be $4$-wise independent random variables over
$\left[  n\right]  $, and let $\mathcal{D}_{i}$\ be the marginal distribution
over $X_{i}$. \ Suppose $K\geq32\sqrt{n}$\ and $\left\Vert \mathcal{D}%
_{i}-\mathcal{D}_{j}\right\Vert \leq1/10$ for all $i,j$. \ Then%
\[
\Pr\left[  \exists i,j:X_{i}=X_{j}\right]  \geq\frac{1}{2}.
\]

\end{numlessthm}


\begin{proof}
Let $Y_{ij}$ be $1$ if $X_{i}=X_{j}$\ and $0$ otherwise, and let
$Y:=\sum_{i<j}Y_{ij}$. \ By Lemma \ref{epsclose}, we have%
\[
\operatorname*{E}\left[  Y\right]  \geq\binom{K}{2}\frac{\left(
1-1/10\right)  ^{2}}{n}\geq900.
\]
The remainder of the proof will involve upper-bounding the second moment
$\operatorname*{E}\left[  Y^{2}\right]  $. \ Let us write%
\[
\operatorname*{E}\left[  Y^{2}\right]  =\sum_{i<j,k<\ell}\operatorname*{E}%
\left[  Y_{ij}Y_{k\ell}\right]  =\tau_{2}+\tau_{3}+\tau_{4},
\]
where $\tau_{N}$ contains the terms in which $N$ distinct indices appear among
$\{i,j,k,l\}$. \ It is easy to see that%
\[
\tau_{2}=\sum_{i<j}\operatorname*{E}\left[  Y_{ij}\right]  =\operatorname*{E}%
\left[  Y\right]
\]
and (by $4$-wise independence) that%
\[
\tau_{4}=\sum_{i<j,k<l~\text{all distinct}}\operatorname*{E}\left[
Y_{ij}\right]  \operatorname*{E}\left[  Y_{k\ell}\right]  \leq
\operatorname*{E}\left[  Y\right]  ^{2}.
\]
So the nontrivial part is to upper-bound $\tau_{3}$. \ Let $p_{i,x}%
:=\Pr\left[  X_{i}=x\right]  $. \ Also, let%
\begin{align*}
r_{x}  &  :=\sum_{i<j<k}p_{i,x}p_{j,x}p_{k,x},\\
s_{x}  &  :=\sum_{i<j}p_{i,x}p_{j,x},
\end{align*}
and notice that $\sum_{x\in\left[  n\right]  }s_{x}=\operatorname*{E}\left[
Y\right]  $. \ Then%
\begin{align*}
\tau_{3}  &  =6\sum_{i<j<k}\sum_{x\in\left[  n\right]  }p_{i,x}p_{j,x}%
p_{k,x}\\
&  =6\sum_{x\in\left[  n\right]  }r_{x}\\
&  \leq6\sum_{x\in\left[  n\right]  }\sqrt{2s_{x}^{3}}\\
&  \leq6\sqrt{2}\sum_{x\in\left[  n\right]  }\left(  \frac{1}{40}s_{x}%
^{2}+12s_{x}\right) \\
&  \leq6\sqrt{2}\left(  \frac{1}{40}\left(  \sum_{x\in\left[  n\right]  }%
s_{x}\right)  ^{2}+12\sum_{x\in\left[  n\right]  }s_{x}\right) \\
&  =6\sqrt{2}\left(  \frac{\operatorname*{E}\left[  Y\right]  ^{2}}%
{40}+12\operatorname*{E}\left[  Y\right]  \right)  .
\end{align*}
Here the third line follows from Lemma \ref{sortlem}, and the fourth line
follows from the basic calculus fact that $s^{3/2}\leq\frac{1}{40}s^{2}%
+12s$\ for all nonnegative $s$. Hence%
\begin{align*}
\Pr\left[  Y=0\right]   &  \leq\Pr\left[  \left\vert Y-\operatorname*{E}%
\left[  Y\right]  \right\vert \geq\operatorname*{E}\left[  Y\right]  \right]
\\
&  =\Pr\left[  \left(  Y-\operatorname*{E}\left[  Y\right]  \right)  ^{2}%
\geq\operatorname*{E}\left[  Y\right]  ^{2}\right] \\
&  \leq\frac{\operatorname*{Var}\left[  Y\right]  }{\operatorname*{E}\left[
Y\right]  ^{2}}\\
&  =\frac{\operatorname*{E}\left[  Y^{2}\right]  -\operatorname*{E}\left[
Y\right]  ^{2}}{\operatorname*{E}\left[  Y\right]  ^{2}}\\
&  =\frac{\tau_{2}+\tau_{3}+\tau_{4}-\operatorname*{E}\left[  Y\right]  ^{2}%
}{\operatorname*{E}\left[  Y\right]  ^{2}}\\
&  \leq\frac{\operatorname*{E}\left[  Y\right]  +6\sqrt{2}\left(
\operatorname*{E}\left[  Y\right]  ^{2}/40+12\operatorname*{E}\left[
Y\right]  \right)  +\operatorname*{E}\left[  Y\right]  ^{2}-\operatorname*{E}%
\left[  Y\right]  ^{2}}{\operatorname*{E}\left[  Y\right]  ^{2}}\\
&  =\frac{1+72\sqrt{2}}{\operatorname*{E}\left[  Y\right]  }+\frac{6\sqrt{2}%
}{40}\\
&  \leq\frac{1}{2}.
\end{align*}
Finally,%
\[
\Pr\left[  \exists i,j:X_{i}=X_{j}\right]  =1-\Pr\left[  Y=0\right]  \geq
\frac{1}{2}%
\]
and we are done.
\end{proof}

\subsection{General Observations\label{REMARKS}}

We conclude this subsection by making three general observations about Theorem
\ref{3satthm}.

First, we strongly believe that our protocol can be improved to one involving
two provers, one of whom sends $O\left(  \log m\right)  $\ qubits and the
other of whom sends $O(\sqrt{m}\operatorname*{polylog}m)$ qubits.
\ Specifically, if all but one of the witnesses in our current protocol are
entangled with one another, in a way that breaks the protocol's soundness, we
believe Arthur should be able to use the remaining witness to detect this.
\ This is a problem we leave to future work.

Second, our protocol made essential use of the PCP Theorem, in the strong
version proved by Dinur \cite{dinur}. \ One might wonder whether Theorem
\ref{3satthm}\ could also be proved in a \textquotedblleft
black-box\textquotedblright\ fashion, without exploiting anything about the
structure of \textsc{3Sat}. \ The following simple result shows that the
answer is no---and that indeed, in the black-box setting, there is essentially
no savings at all over the classical witness size.

\begin{theorem}
\label{nosavings}Let $f:\left\{  0,1\right\}  ^{n}\rightarrow\left\{
0,1\right\}  $ be a black-box function. \ Then any $\mathsf{QMA}^{f}\left(
k\right)  $\ protocol to convince Arthur that there exists an $x$ such that
$f\left(  x\right)  =1$, with soundness gap $\Omega\left(
1/\operatorname*{poly}\left(  n\right)  \right)  $, must involve $n-O\left(
\log n\right)  $ qubits sent by the Merlins.
\end{theorem}

\begin{proof}
[Proof Sketch]Assume without loss of generality that either $f$\ is
identically zero, or else there exists a unique \textquotedblleft marked
item\textquotedblright\ $x^{\ast}$\ such that $f\left(  x^{\ast}\right)  =1$.
\ Suppose it were possible to convince Arthur that $x^{\ast}$ exists by giving
him unentangled witnesses $\left\vert \varphi_{1}\right\rangle ,\ldots
,\left\vert \varphi_{K}\right\rangle $\ with\ $Q$\ qubits in total. \ Then
given these witnesses, Arthur's verification algorithm must query $f\left(
x^{\ast}\right)  $\ at some time step with non-negligible probability
$\beta=\Omega\left(  1/\operatorname*{poly}\left(  n\right)  \right)  $. \ For
otherwise, by the hybrid argument\ of Bennett, Bernstein, Brassard, and
Vazirani\ \cite{bbbv}, Arthur's verification algorithm would not have
$\Omega\left(  1/\operatorname*{poly}\left(  n\right)  \right)  $\ soundness
(i.e., Arthur would fail to detect a change in $f\left(  x^{\ast}\right)
$\ from $1$ to $0$). \ But this means that Arthur's algorithm can be modified
to one that uses no witnesses, and that \textit{finds} $x^{\ast}$ with
probability at least $2^{-Q}\beta/T$.\ \ For Arthur can simply replace
$\left\vert \varphi_{1}\right\rangle ,\ldots,\left\vert \varphi_{K}%
\right\rangle $ by the $Q$-qubit maximally mixed state, then measure at a
random time step to find which $x$\ is being queried. On the other hand, we
know from the result of Bennett et al.\ \cite{bbbv} mentioned previously that
if $x^{\ast}$ is uniformly random, then after $T$ queries, Arthur can have
found $x^{\ast}$\ with probability at most $4T^{2}/2^{n}$. \ Solving
$2^{-Q}\beta/T\leq4T^{2}/2^{n}$\ for $Q$, we find that%
\[
Q\geq\log\frac{\beta2^{n}}{4T^{3}}\geq n-O\left(  \log n\right)  .
\]

\end{proof}

Third, notice that our protocol does not let Arthur \textit{find} a satisfying
assignment for $\varphi$; it only convinces him that such an assignment
exists. \ If there were a way to modify our protocol to let Arthur recover an
assignment, this would have a spectacular consequence for quantum algorithms.
\ Namely, by running Arthur's verification procedure with the $\widetilde
{O}\left(  \sqrt{m}\right)  $-qubit\ maximally mixed state in place of the
witnesses, we could find a satisfying assignment for $\varphi$ with
probability $2^{-\widetilde{O}\left(  \sqrt{m}\right)  }$, with no help from
any Merlins. \ But this would yield a $2^{\widetilde{O}\left(  \sqrt
{m}\right)  }$-time quantum algorithm for \textsc{3Sat}---and in particular, a
$2^{\widetilde{O}\left(  \sqrt{n}\right)  }$-time\ algorithm in the
\textquotedblleft critical regime\textquotedblright\ $m=O\left(  n\right)  $!

\section{Weak Additivity Implies Amplification\label{AMP}}

In this section we show how to amplify any $\mathsf{QMA}\left(  k\right)
$\ protocol to exponentially small error, and to simulate $k$ provers with
two, assuming a weak version of the Additivity Conjecture.

\subsection{Entanglement of Formation\label{EF}}

The analysis of our amplification protocol will involve showing that Arthur
cannot create \textquotedblleft too much\textquotedblright\ entanglement
during his verification procedure. \ To make this precise, we need some way to
measure the entanglement of mixed states. \ Fortunately, this is one of the
most studied topics in all of quantum information theory. \ One particular
entanglement measure---the \textit{entanglement of formation} $E_{F}$\ defined
by Bennett et al. \cite{bdsw}---will be particularly useful for us.

To define $E_{F}$\ we first need some other concepts. \ Given a mixed state
$\sigma$, the \textit{von Neumann entropy} of $\sigma$\ is $S\left(
\sigma\right)  :=H\left(  \left\{  \lambda_{i}\right\}  \right)  $, where
$H\left(  \left\{  p_{i}\right\}  \right)  =-\sum_{i}p_{i}\log_{2}p_{i}$\ is
the usual entropy function and $\left\{  \lambda_{i}\right\}  $\ are the
eigenvalues of $\sigma$. \ Given a bipartite pure state $\left\vert \psi
^{AB}\right\rangle $, let $\sigma^{A}$\ and $\sigma^{B}$\ be the reduced
states of the $A$ and $B$ registers respectively. \ Then it is not hard to
show that $S\left(  \sigma^{A}\right)  =S\left(  \sigma^{B}\right)  $. \ We
call this quantity the \textit{entanglement entropy} of $\left\vert \psi
^{AB}\right\rangle $, or $E(\left\vert \psi^{AB}\right\rangle )$. \ We can
then define $E_{F}(\rho^{AB})$\ as a weighted average of entanglement entropy,
minimized over all purifications of $\rho^{AB}$:

\begin{definition}
Given a bipartite state $\rho^{AB}$, the entanglement of formation $E_{F}%
(\rho^{AB})$\ is the minimum of $\sum_{i}p_{i}E\left(  \left\vert \psi
_{i}\right\rangle \right)  $\ over all decompositions $\rho^{AB}=\sum_{i}%
p_{i}\left\vert \psi_{i}\right\rangle \left\langle \psi_{i}\right\vert $.
\end{definition}

Intuitively, $E_{F}$\ measures the minimum number of entangled pairs\ $\frac
{1}{\sqrt{2}}\left(  \left\vert 00\right\rangle +\left\vert 11\right\rangle
\right)  $ that are needed to prepare $\rho^{AB}$.

Almost by definition, $E_{F}$\ satisfies \textit{convexity}: for all
$\rho^{AB}$\ and $\sigma^{AB}$,%
\[
E_{F}\left(  p\rho^{AB}+\left(  1-p\right)  \sigma^{AB}\right)  \leq
pE_{F}\left(  \rho^{AB}\right)  +\left(  1-p\right)  E_{F}\left(  \sigma
^{AB}\right)  .
\]
It is also easy to see that $E_{F}\left(  \rho^{AB}\right)  =0$\ if and only
if $\rho^{AB}$ is separable. \ In this paper, we will need two further
properties of $E_{F}$. \ The first property is what we called
\textquotedblleft faithfulness\textquotedblright\ in Section \ref{AMPINT}.

\begin{lemma}
\label{eflem}Suppose $E_{F}(\rho^{AB})\leq\varepsilon$. \ Then there exists a
separable state that is $\sqrt{2\varepsilon}$-close to $\rho^{AB}$\ in trace distance.
\end{lemma}

\begin{proof}
Let $S\left(  \rho||\sigma\right)  $\ be the \textit{quantum relative entropy}
between mixed states $\rho$ and $\sigma$ (see Nielsen and Chuang
\cite{nc}\ for a definition). \ Then Vedral and Plenio \cite{vp} showed that
\[
E_{F}(\rho^{AB})\geq\min S\left(  \rho^{AB}||\sigma^{AB}\right)  ,
\]
where the minimum is taken over all separable states $\sigma^{AB}$. \ Also, it
is known (see Klauck et al.\ \cite{kntz} and Ohya and Petz \cite{ohyapetz}%
)\ that%
\[
S(\rho^{AB}||\sigma^{AB})\geq\frac{1}{2}\left\Vert \rho^{AB}-\sigma
^{AB}\right\Vert _{\operatorname*{tr}}^{2}.
\]
Putting these results together, if $E_{F}(\rho^{AB})\leq\varepsilon$\ then
there exists a separable state $\sigma^{AB}$ such that $S\left(  \rho
^{AB}||\sigma^{AB}\right)  \leq\varepsilon$, and hence $\left\Vert \rho
^{AB}-\sigma^{AB}\right\Vert _{\operatorname*{tr}}\leq\sqrt{2\varepsilon}$.
\end{proof}

The second property is that, if we start from a separable state, we
cannot increase the value of $E_{F}$\ much by measuring few qubits
and then conditioning on the outcome.

\begin{lemma}
\label{2nlem}Let $\rho^{AB}$ be a separable state, and suppose
$\sigma^{AB}$ is obtained from $\rho^{AB}$\ by applying an arbitrary
entangled measurement on at most $n$ qubits from each register, and
then possibly conditioning on the outcome. \ Then $E_{F}\left(
\sigma^{AB}\right) \leq E_{F}\left( \rho^{AB}\right) +2n$.
\end{lemma}

\begin{proof}
By convexity, we can assume without loss of generality that
$\rho^{AB}$ is a pure state, $\left\vert \psi_{A}\right\rangle
\otimes\left\vert \psi _{B}\right\rangle $. \ So we can write
$\sigma^{A}$ as $\Phi\left(  \left\vert \psi_{A}\right\rangle
\left\langle \psi_{A}\right\vert \right) / \left\Vert \Phi\left(
\left\vert \psi_{A}\right\rangle \left\langle \psi_{A}\right\vert
\right) \right\Vert $, where $\Phi$\ is some non-trace-increasing
operator acting on at most $n$ qubits.
\ Or in the operator-sum representation,%
\[
\sigma^{A}=\frac{\sum_{i=1}^{M}E_{i}\left\vert \psi_{A}\right\rangle
\left\langle \psi_{A}\right\vert E_{i}^{\dagger}}{\operatorname*{Tr}\sum
_{i=1}^{M}E_{i}\left\vert \psi_{A}\right\rangle \left\langle \psi
_{A}\right\vert E_{i}^{\dagger}}%
\]
where $\sum_{i=1}^{2^{2n}}E_{i}^{\dagger}E_{i}\preceq I$ and $M\leq2^{2n}$.
\ We then have%
\begin{align*}
E_{F}\left(  \sigma^{AB}\right)   &  \leq S\left(  \sigma^{A}\right) \\
&  =S\left(  \frac{\sum_{i=1}^{M}E_{i}\left\vert \psi_{A}\right\rangle
\left\langle \psi_{A}\right\vert E_{i}^{\dagger}}{\operatorname*{Tr}\sum
_{i=1}^{M}E_{i}\left\vert \psi_{A}\right\rangle \left\langle \psi
_{A}\right\vert E_{i}^{\dagger}}\right) \\
&  \leq\log_{2}M\\
&  \leq2n
\end{align*}
where the first line follows from the concavity of the von Neumann entropy.
\end{proof}

Given an entanglement measure $E$, we call $E$ \textit{superadditive} if for
any state $\rho^{AA^{\prime},BB^{\prime}}$\ on four registers,%
\[
E(\rho^{AA^{\prime},BB^{\prime}})\geq E\left(  \rho^{AB}\right)
+E(\rho^{A^{\prime}B^{\prime}}).
\]
As mentioned earlier, the analysis of our $\mathsf{QMA}\left(  k\right)
$\ amplification protocol originally relied on the conjecture that $E_{F}$ is
superadditive. \ But in a spectacular recent development, Hastings
\cite{hastings} has shown that this conjecture is false. \ (More precisely,
Hastings shows a failure of additivity for the minimum output entropy of a
quantum channel. \ By a result of Shor \cite{shor:add},\ this is equivalent to
the superadditivity of entanglement of formation.)

However, the violation of additivity found by Hastings is extremely small, and
is perfectly consistent with additivity being true in a weaker or asymptotic
sense. \ To make this precise, we now state a version of the Additivity
Conjecture that suffices for our purposes. \ Call an entanglement measure $E$
\textit{weakly superadditive} if it satisfies the relation%
\[
E\left(  \rho^{A_{1}A_{2}\cdots A_{k},B_{1}B_{2}\cdots B_{k}}\right)
\geq\frac{c}{k}\sum_{i,j=1}^{k}E(\rho^{A_{i}B_{j}}),
\]
for some constant $c$ independent of $k$. \ Weak superadditivity is, in
particular, implied by the following inequality:%
\[
E(\rho^{AA^{\prime},BB^{\prime}})\geq\frac{1}{2}\left[  E\left(  \rho
^{AB}\right)  +E(\rho^{AB^{\prime}})+E(\rho^{A^{\prime}B})+E(\rho^{A^{\prime
}B^{\prime}})\right]
\]
which in turn is implied by ordinary superadditivity. \ Then we conjecture the following:

\begin{conjecture}
[Weak Additivity Conjecture]\label{weakaddconj}$E_{F}$ is weakly superadditive.
\end{conjecture}

As a side note, $E_{F}$\ badly violates the so-called \textit{monogamy
inequality} $E(\rho^{A,BB^{\prime}})\geq E\left(  \rho^{AB}\right)
+E(\rho^{AB^{\prime}})$.\footnote{Note that the monogamy inequality implies
superadditivity, via%
\begin{align*}
E(\rho^{AA^{\prime},BB^{\prime}}) &  \geq E(\rho^{AA^{\prime},B}%
)+E(\rho^{AA^{\prime},B^{\prime}})\\
&  \geq E(\rho^{AB})+E(\rho^{A^{\prime}B})+E(\rho^{AB^{\prime}})+E(\rho
^{A^{\prime}B^{\prime}})\\
&  \geq E(\rho^{AB})+E(\rho^{A^{\prime}B^{\prime}}).
\end{align*}
} \ As an example, consider again the maximally antisymmetric state%
\[
\left\vert \psi\right\rangle =\frac{1}{\sqrt{N!}}\sum_{\sigma\in S_{N}}\left(
-1\right)  ^{\operatorname*{sgn}\left(  \sigma\right)  }\left\vert
\sigma\left(  1\right)  \right\rangle \cdots\left\vert \sigma\left(  N\right)
\right\rangle .
\]
The entanglement of formation between the first register of
$\left\vert \psi\right\rangle $\ and the remaining $N-1$\ registers
is at most $\log N$. \ Yet the entanglement of formation between\
the first register and any \textit{one} other register can be shown
to be $\Omega\left(  1\right)  $. \ Note that, had $E_{F}$ satisfied
the monogamy inequality, we would have been able to use it to show
$\mathsf{QMA}\left( 2\right) = \mathsf{QMA}$, using essentially the
same argument as in footnote \ref{SCHMIDT}.

A different entanglement measure---the \textit{squashed entanglement} $E_{sq}%
$\ of Christandl and Winter \cite{squashed}---is known to satisfy both
superadditivity and the stronger monogamy inequality. \ The trouble with
$E_{sq}$ is that it badly violates the analogue of Lemma \ref{eflem}: there
exist $N\times N$-dimensional bipartite states $\rho^{AB}$\ such that
$E_{sq}(\rho^{AB})=O\left(  \frac{\log N}{N}\right)  $, yet $\rho^{AB}$\ has
trace distance $\Omega\left(  1\right)  $\ to any separable state. \ The
example that shows this is once again the maximally antisymmetric state, which
seems like the \textquotedblleft universal counterexample\textquotedblright%
\ of entanglement theory! \ This is why we cannot use squashed entanglement in
this paper, and must instead use entanglement of formation.

\subsection{The Two-Prover Case\label{2PROVER}}

We now show that the Weak Additivity Conjecture implies the
$\mathsf{QMA}\left( 2\right)  $\ Amplification Conjecture.

\begin{theorem}
\label{ampthm}Assume the Weak Additivity Conjecture. \ Then $\mathsf{QMA}%
\left(  2,a,b\right)  =\mathsf{QMA}\left(  2,2^{-p\left(  n\right)
},1-2^{-p\left(  n\right)  }\right)  $ for all $b-a=\Omega\left(
1/\operatorname*{poly}\left(  n\right)  \right)  $\ and all polynomials $p$.
\end{theorem}

\begin{proof}
Let $L$ be a language in $\mathsf{QMA}\left(  2,a,b\right)  $; then
we need to show $L\in\mathsf{QMA}\left(  2,2^{-p\left(  n\right)
},1-2^{-p\left( n\right)  }\right)  $. \ Let $Q$ be Arthur's
verification algorithm in the original $\mathsf{QMA}\left(
2,a,b\right)  $\ protocol, and let the original Merlins' messages
have $r\left(  n\right)  $\ qubits each for some polynomial $r$. \
Also, let $T\left(  n\right)  $\ be a number of repetitions of $Q$
that suffices to amplify it to error probability $2^{-p\left(
n\right)  }$, assuming no entanglement among Merlin$_{A}$'s or
Merlin$_{B}$'s registers. \ By a standard Chernoff bound, we can
take$\ T\left(  n\right)  :=C\cdot p\left(  n\right)  /\left(
b-a\right)  ^{2}$ for some constant $C$.

Our amplified protocol is the following.

\begin{itemize}
\item[(1)] Arthur asks Merlin$_{A}$ and Merlin$_{B}$ to supply $q\left(
n\right)  $\ copies each of their respective witnesses, where $q\left(
n\right)  :=C^{\prime}\cdot T\left(  n\right)  r\left(  n\right)  /\left(
b-a\right)  ^{2}$ for some constant $C^{\prime}$. \ Denote by $\rho
^{A_{1}A_{2}\cdots A_{q\left(  n\right)  }}$\ and $\rho^{B_{1}B_{2}\cdots
B_{q\left(  n\right)  }}$\ the states on $q\left(  n\right)  r\left(
n\right)  $\ qubits that Arthur actually receives.

\item[(2)] For all $t:=1$\ to $T\left(  n\right)  $, Arthur chooses registers
$A_{j}$\ and $B_{k}$\ uniformly and independently from among those not already
chosen, and runs $Q$ on the state $\rho^{A_{j}B_{k}}$.

\item[(3)] Arthur accepts if at least $\frac{a+b}{2}T\left(  n\right)  $\ of
the $T\left(  n\right)  $\ invocations of $Q$ accepted,\ and rejects otherwise.
\end{itemize}

We need to show two things about this protocol, completeness and soundness.

\textbf{Completeness:}\ If the Merlins are honest, they can simply send
$\left\vert \psi_{A}\right\rangle ^{\otimes q\left(  n\right)  }$\ and
$\left\vert \psi_{B}\right\rangle ^{\otimes q\left(  n\right)  }%
$\ respectively, where $\left\vert \psi_{A}\right\rangle \otimes\left\vert
\psi_{B}\right\rangle $\ is a witness that $Q$\ accepts with probability at
least $b$. \ Then by assumption, Arthur will accept with probability at least
$1-2^{-p\left(  n\right)  }$.

\textbf{Soundness:} As usual, this is the interesting part. \ Our central
claim is the following: \textit{At every one of the }$T\left(  n\right)
$\textit{ iterations, Arthur can be considered to be running }$Q$\textit{\ on
a bipartite state }$\rho^{AB}$\textit{\ that is }$\varepsilon$\textit{-close
to a separable state, where }$\varepsilon=O\left(  \sqrt{T\left(  n\right)
r\left(  n\right)  /q\left(  n\right)  }\right)  $\textit{.}

Let us first see why soundness follows from the above claim. \ Suppose
$x\notin L$. \ Then $Q$ accepts every separable state with probability at most
$a$. \ By Proposition \ref{closeprop}, then, $Q$ also accepts every state that
is $\varepsilon$-close\ to separable with probability at most\ $a+\varepsilon
$. \ But%
\[
\varepsilon=O\left(  \sqrt{\frac{T\left(  n\right)  r\left(  n\right)
}{q\left(  n\right)  }}\right)  \leq\frac{b-a}{4},
\]
provided we chose a sufficiently large constant $C^{\prime}$\ when defining
$q\left(  n\right)  $. \ So every invocation of $Q$ accepts with probability
at most $a+\frac{b-a}{4}$. \ Therefore, provided we choose a sufficiently
large constant $C$ when defining $T\left(  n\right)  $, Arthur will accept
with probability at most $2^{-p\left(  n\right)  }$ by a Chernoff bound.

We now prove the claim. \ By Lemma \ref{2nlem}, the entanglement of formation
between Merlin$_{A}$'s registers and Merlin$_{B}$'s registers can be at most
$2r\left(  n\right)  $ after the first iteration, at most $4r\left(  n\right)
$\ after the second iteration, and so on. \ Hence%
\[
E_{F}\left(  \rho^{A_{1}A_{2}\cdots A_{q\left(  n\right)  },B_{1}B_{2}\cdots
B_{q\left(  n\right)  }}\right)  \leq2T\left(  n\right)  r\left(  n\right)
\]
throughout the protocol. \ Also, let $S_{A}$ and $S_{B}$ be the sets of
$A$-registers and $B$-registers respectively that Arthur has not yet chosen.
\ Then $\left\vert S_{A}\right\vert =\left\vert S_{B}\right\vert =q\left(
n\right)  -T\left(  n\right)  $. \ Assuming the Weak Additivity Conjecture, we
therefore have%
\begin{align*}
\sum_{A_{j}\in S_{A},B_{k}\in S_{B}}E_{F}\left(  \rho^{A_{j}B_{k}}\right)   &
=O\left(  \left(  q\left(  n\right)  -T\left(  n\right)  \right)  E_{F}\left(
\rho^{A_{1}A_{2}\cdots A_{q\left(  n\right)  },B_{1}B_{2}\cdots B_{q\left(
n\right)  }}\right)  \right)  \\
&  =O\left(  T\left(  n\right)  r\left(  n\right)  \left(  q\left(  n\right)
-T\left(  n\right)  \right)  \right)  .
\end{align*}
So if we define%
\[
\sigma:=\frac{1}{\left\vert S_{A}\right\vert \left\vert S_{B}\right\vert }%
\sum_{A_{j}\in S_{A},B_{k}\in S_{B}}\rho^{A_{j}B_{k}},
\]
then the convexity of $E_{F}$ implies that%
\begin{align*}
E_{F}\left(  \sigma\right)   &  \leq\frac{1}{\left\vert S_{A}\right\vert
\left\vert S_{B}\right\vert }\sum_{A_{j}\in S_{A},B_{k}\in S_{B}}E_{F}\left(
\rho^{A_{j}B_{k}}\right)  \\
&  =O\left(  \frac{T\left(  n\right)  r\left(  n\right)  }{q\left(  n\right)
-T\left(  n\right)  }\right)  \\
&  =O\left(  \frac{T\left(  n\right)  r\left(  n\right)  }{q\left(  n\right)
}\right)  ,
\end{align*}
using the fact that $T\left(  n\right)  \leq q\left(  n\right)  /2$. \ By
Lemma \ref{eflem}, this means that $\sigma$\ is $O\left(  \sqrt{T\left(
n\right)  r\left(  n\right)  /q\left(  n\right)  }\right)  $-close\ to a
separable state, as claimed.
\end{proof}

\subsection{The $k$-Prover Case\label{KPROVER}}

In this section we show unconditionally that \textit{any }$\mathsf{QMA}\left(
k\right)  $\textit{ protocol with constant soundness can be simulated by a
}$\mathsf{QMA}\left(  2\right)  $\textit{ protocol with soundness }%
$\Omega\left(  1/k\right)  $. \ Combined with Theorem \ref{ampthm}, this will
imply that assuming the Weak Additivity Conjecture, for all $k\geq2$\ we have%
\[
\mathsf{QMA}\left(  k,1/3,2/3\right)  \subseteq\mathsf{QMA}\left(
2,a,b\right)  \subseteq\mathsf{QMA}\left(  2,1/3,2/3\right)
\]
for some $a,b$\ with $b-a=\Omega\left(  1/k\right)  $, and hence
$\mathsf{QMA}\left(  k\right)  =\mathsf{QMA}\left(  2\right)  $. \ Note that
Kobayashi et al. (personal communication) have independently shown that
amplification of $\mathsf{QMA}\left(  2\right)  $\ protocols implies
$\mathsf{QMA}\left(  k\right)  =\mathsf{QMA}\left(  2\right)  $\ for all
$k\geq2$. \ (Their earlier result only showed that amplification of
$\mathsf{QMA}\left(  k\right)  $\ protocols implies $\mathsf{QMA}\left(
k\right)  =\mathsf{QMA}\left(  2\right)  $\ for all $k\geq2$, which was not
strong enough for our purposes.)

\begin{theorem}
\label{ktotwo}$\mathsf{QMA}\left(  k,a,b\right)  \subseteq\mathsf{QMA}\left(
2,1-\frac{\left(  b-a\right)  ^{2}}{8k},1-2^{-n}\right)  $.
\end{theorem}

\begin{proof}
We will show that for all $k$ and all $\delta=\Omega\left(
1/\operatorname*{poly}\left(  n\right)  \right)  $,%
\[
\mathsf{QMA}\left(  k,1-\delta,1-2^{-n}\right)  \subseteq\mathsf{QMA}\left(
2,1-\frac{\delta^{2}}{8k},1-2^{-n}\right)  .
\]
This will suffice to prove the theorem, since Lemma
\ref{onesidedamp} implies that for all $k$ and all $a,b$, we have
$\mathsf{QMA}\left(  k,a,b\right) \subseteq\mathsf{QMA}\left(
k,1-\left(  b-a\right)  ,1-2^{-n}\right)  $.

Our protocol is as follows. \ Merlin$_{A}$ and Merlin$_{B}$\ send
$k$-partite
states $\rho^{A_{1}A_{2}\cdots A_{k}}$\ and $\rho^{B_{1}B_{2}\cdots B_{k}}%
$\ respectively. \ Given these states, Arthur performs one of the following
two tests, each with probability $1/2$:

\begin{enumerate}
\item[(1)] Choose $i\in\left[  k\right]  $\ uniformly at random, perform a
swap test between $\rho^{A_{i}}$\ and $\rho^{B_{i}}$, and accept if and only
if the swap test accepts.

\item[(2)] Simulate the $\mathsf{QMA}\left(  k,1-\delta,1-2^{-n}\right)
$\ protocol, using $\rho^{A_{1}A_{2}\cdots A_{k}}$\ in place of the $k$
witness registers.
\end{enumerate}

We first show completeness of the above protocol. \ If the Merlins
are honest, they can both simply send $k$ unentangled accepting
witnesses for the $\mathsf{QMA}\left(  k\right)  $ protocol being
simulated. \ In that case step (1) accepts with probability $1$,
while step (2) accepts with probability at least $1-2^{-n}$.

We now show soundness. \ Suppose any set of unentangled witnesses\
causes the $\mathsf{QMA}\left(  k\right)  $ protocol to reject with
probability at least $\delta$. \ Then we need to show that any pair
of
witnesses $\rho^{A_{1}A_{2}\cdots A_{k}}$\ and $\rho^{B_{1}B_{2}\cdots B_{k}}%
$\ causes the $\mathsf{QMA}\left(  2\right)  $\ protocol to reject
with probability at least $\frac{\delta^{2}}{8k}$. \ We consider two
cases.

First suppose $\rho^{A_{1}A_{2}\cdots A_{k}}$\ is
$\varepsilon$-close\ in trace distance to some separable pure state
$\left\vert \Psi\right\rangle $. \ Then by Proposition
\ref{closeprop}, step (2) rejects with probability at least
$\delta-\varepsilon$.

Next suppose $\rho^{A_{1}A_{2}\cdots A_{k}}$\ is $\varepsilon$-far
in trace distance from any separable pure state. \ Then by
Proposition \ref{fidelityprop}, we have $\left\langle \Psi|\rho^{A_{1}%
A_{2}\cdots A_{k}}|\Psi\right\rangle <1-\varepsilon^{2}$\ for all separable
pure states $\left\vert \Psi\right\rangle $. \ So taking the contrapositive of
Proposition \ref{closetopure}, for all pure states $\left\vert \psi
_{1}\right\rangle ,\ldots,\left\vert \psi_{k}\right\rangle $ we have\
\[
\sum_{i=1}^{k}\left(  1-\left\langle \psi_{i}|\rho^{A_{i}}|\psi_{i}%
\right\rangle \right)  >\varepsilon^{2}.
\]
Hence step (1) rejects with probability greater than $\frac{\varepsilon^{2}%
}{2k}$ by Proposition \ref{swaptest}. Setting $\varepsilon=3\delta/4$, we thus
find that the protocol rejects with probability at least $\frac{1}{2}%
\max\left\{  \frac{\delta}{4},\frac{\left(  3\delta/4\right)  ^{2}}%
{2k}\right\}  \geq\frac{\delta^{2}}{8k}$.
\end{proof}

Combining Theorem \ref{ktotwo} with Theorem \ref{ampthm} now yields the following:

\begin{corollary}
\label{ampcor}The Weak Additivity Conjecture implies the Collapse Conjecture,
that $\mathsf{QMA}\left(  k\right)  =\mathsf{QMA}\left(  2\right)  $ for all
$k\geq2$.
\end{corollary}

\subsection{\label{SYMQMA}Symmetric $\mathsf{QMA}\left(  k\right)  $}

Define the complexity class $\mathsf{SymQMA}\left(  k,a,b\right)  $\ the same
way as $\mathsf{QMA}\left(  k,a,b\right)  $, except that now we are promised
that the $k$ witnesses are all identical (in both the completeness and
soundness cases). \ We saw in Section \ref{SYMMETRIC} that symmetric
$\mathsf{QMA}\left(  k\right)  $\ protocols are sometimes easier to analyze
than non-symmetric ones. \ However, we will now show that assuming the Weak
Additivity Conjecture, $\mathsf{QMA}\left(  k\right)  $\ and $\mathsf{SymQMA}%
\left(  k\right)  $\ are actually equivalent.\footnote{This does not
mean that proving the Weak Additivity Conjecture would supersede the
analysis of the non-symmetric case in Section \ref{3SAT}. \ Our
result will only show that,
assuming the Weak Additivity Conjecture, the witness sizes\ in $\mathsf{QMA}%
$\ and $\mathsf{SymQMA}$\ protocols are \textit{polynomially} related, which
is vacuous in the context of our \textsc{3Sat}\ protocol.}

The first step is to show they are (unconditionally) equivalent up to a loss
in error bounds.

\begin{lemma}
\label{symlem}$\mathsf{QMA}\left(  k,a,b\right)  \subseteq\mathsf{SymQMA}%
\left(  k,a,b\right)  \subseteq\mathsf{QMA}\left(  k,1-\frac{\left(
b-a\right)  ^{2}}{8k},1-2^{-n}\right)  $.
\end{lemma}

\begin{proof}
For the first containment, have each Merlin in the $\mathsf{SymQMA}$\ protocol
send $k$ witnesses (for a total of $k^{2}$\ witnesses). \ Then simulate the
$\mathsf{QMA}$\ protocol by using the $i^{th}$\ witness from the $i^{th}%
$\ Merlin for all $i\in\left[  k\right]  $. For the second containment, first
observe that%
\[
\mathsf{SymQMA}\left(  k,a,b\right)  \subseteq\mathsf{SymQMA}\left(
k,1-\left(  b-a\right)  ,1-2^{-n}\right)  ,
\]
completely analogously to Lemma \ref{onesidedamp}. \ Let $\delta=b-a$. \ Then
to simulate a $\mathsf{SymQMA}\left(  k,1-\delta,1-2^{-n}\right)  $\ protocol
without the symmetry promise we do the following. \ Let $\left\vert
\varphi_{i}\right\rangle $\ be the witness sent by the $i^{th}$\ Merlin. \ Then

\begin{itemize}
\item With $1/2$ probability, Arthur performs a swap test between $\left\vert
\varphi_{1}\right\rangle $\ and a random other witness,\ and accepts if and
only if the swap test accepts.

\item With $1/2$ probability, Arthur runs the $\mathsf{SymQMA}$\ protocol as
if the witnesses were identical.
\end{itemize}

In the completeness case, it is clear that Arthur accepts with probability
greater than $1-2^{-n}$. To show soundness we consider two cases, just like in
Theorem \ref{ktotwo}. \ Let $\left\vert \Phi\right\rangle =\left\vert
\varphi_{1}\right\rangle \otimes\cdots\otimes\left\vert \varphi_{k}%
\right\rangle $. \ First suppose $\left\vert \Phi\right\rangle $\ is
$\varepsilon$-close\ in trace distance to $\left\vert \varphi_{1}\right\rangle
^{\otimes k}$. \ Then by Proposition \ref{closeprop}, when he runs the
$\mathsf{SymQMA}$\ protocol\ Arthur will reject with probability at least
$\delta-\varepsilon$. Next suppose $\left\vert \Phi\right\rangle $\ is
$\varepsilon$-far from $\left\vert \varphi_{1}\right\rangle ^{\otimes k}$.
\ Then $| \left\langle \varphi_{1}\right\vert ^{\otimes k}\left\vert
\Phi\right\rangle | ^{2}<1-\varepsilon^{2}$\ by Proposition \ref{fidelityprop}%
. \ So by Proposition \ref{closetopure}, we have%
\[
\sum_{i=1}^{k}\left(  1-\left\vert \left\langle \varphi_{1}|\varphi
_{i}\right\rangle \right\vert ^{2}\right)  >\varepsilon^{2}.
\]
Hence when Arthur performs a swap test, he rejects with probability
greater than $\frac{\varepsilon^{2}}{2k}$.

Setting $\varepsilon:=3\delta/4$, we thus
find that the protocol rejects with probability at least $\frac{\delta^{2}%
}{8k}$.
\end{proof}

Combining Lemma \ref{symlem} with Theorem \ref{ktotwo},\ we immediately get
the following.

\begin{theorem}
The $\mathsf{QMA}\left(  2\right)  $\ amplification conjecture implies
$\mathsf{SymQMA}\left(  k\right)  =\mathsf{QMA}\left(  k\right)
=\mathsf{QMA}\left(  2\right)  $\ for all $k\geq2$.
\end{theorem}

\section{Nonexistence of Perfect Disentanglers\label{GDM}}

In this section, we will speak interchangeably about Hilbert spaces and spaces
of density operators over those Hilbert spaces.

\begin{definition}
Let $\mathcal{H}$ and $\mathcal{K}$ be two finite-dimensional Hilbert spaces.
\ Then given a superoperator $\Phi:\mathcal{H}\rightarrow\mathcal{K}%
\otimes\mathcal{K}$, we say $\Phi$\ is an $\left(  \varepsilon,\delta\right)
$-disentangler if

\begin{enumerate}
\item[(i)] $\Phi(\rho)$ is $\varepsilon$-close to a separable state for every
$\rho$, and

\item[(ii)] for every separable state $\sigma$, there exists a $\rho$ such
that $\Phi\left(  \rho\right)  $\ is $\delta$-close to $\sigma$.
\end{enumerate}
\end{definition}

As pointed out in Section \ref{GDMINT}, if for sufficiently small constants
$\varepsilon,\delta$\ there exists an $\left(  \varepsilon,\delta\right)
$-disentangler with $\log\dim\mathcal{H}=\operatorname*{poly}\left(  \log
\dim\mathcal{K}\right)  $---and if, moreover, that disentangler can be
implemented in quantum polynomial time---then $\mathsf{QMA}\left(  2\right)
=\mathsf{QMA}$.

Watrous (personal communication) has proposed the following fundamental conjecture.

\begin{conjecture}
[Watrous]\label{gdmconj}For all constants $\varepsilon,\delta<1$, any $\left(
\varepsilon,\delta\right)  $-disentangler requires $\dim\mathcal{H}%
=2^{\Omega\left(  \dim\mathcal{K}\right)  }$.
\end{conjecture}

A proof of Conjecture \ref{gdmconj}\ would be an important piece of formal
evidence that $\mathsf{QMA}\left(  2\right)  \neq\mathsf{QMA}$, and might even
lead to a \textquotedblleft quantum oracle separation\textquotedblright\ (as
defined by Aaronson and Kuperberg \cite{ak}) between the two classes.

Here we show that, at least in the case $\varepsilon=\delta=0$, no
disentangler exists in \textit{any} finite dimension. \ The counterexamples in
Section \ref{GDMINT} imply that this result\ would be false if we let either
$\varepsilon$\ or $\delta$\ be nonzero.

\begin{theorem}
\label{gdmthm}Let $\Phi:\mathcal{H}\rightarrow\mathcal{K}\otimes\mathcal{K}$
be any superoperator whose image is the set of separable states. \ Then
$\dim\mathcal{K}\geq2$\ implies $\dim\mathcal{H}=\infty$.
\end{theorem}

\begin{proof}
For any pure state $\left\vert \alpha\right\rangle \in\mathcal{K}$, by
assumption there exists a state $\rho_{\alpha}$ such that $\Phi(\rho_{\alpha
})=\left\vert \alpha\right\rangle \left\langle \alpha\right\vert
\otimes\left\vert \alpha\right\rangle \left\langle \alpha\right\vert $. \ By
convexity, we can assume $\rho_{\alpha}=\left\vert \phi_{\alpha}\right\rangle
\left\langle \phi_{\alpha}\right\vert $ is pure. \ Also, suppose
$\dim\mathcal{H}$ is finite. \ Then $\Phi$ admits an
operator-sum\ representation $\Phi(\rho)=\sum_{i=1}^{k}E_{i}\rho
E_{i}^{\dagger}$\ where $\sum_{i=1}^{k}E_{i}^{\dagger}E_{i}=I$. \ We then
have
\[
\Phi(\left\vert \phi_{\alpha}\right\rangle \left\langle \phi_{\alpha
}\right\vert )=\sum_{i=1}^{k}E_{i}\left\vert \phi_{\alpha}\right\rangle
\left\langle \phi_{\alpha}\right\vert E_{i}^{\dagger}=\left\vert
\alpha\right\rangle \left\langle \alpha\right\vert \otimes\left\vert
\alpha\right\rangle \left\langle \alpha\right\vert .
\]
So again by convexity, we find that $E_{i}\left\vert \phi_{\alpha
}\right\rangle $ must be a multiple of $\left\vert
\alpha\right\rangle \left\vert \alpha\right\rangle $ for all $i$\
and $\alpha$; that is, there exist constants $c_{\alpha,i}$ such
that $E_{i}\left\vert \phi_{\alpha }\right\rangle
=c_{\alpha,i}\left\vert \alpha\right\rangle \left\vert
\alpha\right\rangle $.

Now let $\left\vert \alpha\right\rangle ,\left\vert
\beta\right\rangle $ be any two pure states in$\ \mathcal{K}$ with
$\left\vert \alpha\right\rangle \neq\left\vert \beta\right\rangle $.
\ Also let $\left\vert \psi\right\rangle =a\left\vert
\phi_{\alpha}\right\rangle +b\left\vert \phi_{\beta}\right\rangle $
for some nonzero real numbers $a,b$.
\ Then%
\begin{align*}
\Phi(\left\vert \psi\right\rangle \left\langle \psi\right\vert )  &
=a^{2}\Phi(\left\vert \phi_{\alpha}\right\rangle \left\langle \phi_{\alpha
}\right\vert )+b^{2}\Phi(\left\vert \phi_{\beta}\right\rangle \left\langle
\phi_{\beta}\right\vert )+ab\Phi(\left\vert \phi_{\alpha}\right\rangle
\left\langle \phi_{\beta}\right\vert )+ab\Phi(\left\vert \phi_{\beta
}\right\rangle \left\langle \phi_{\alpha}\right\vert )\\
&  =a^{2}\left\vert \alpha\right\rangle \left\langle \alpha\right\vert
\otimes\left\vert \alpha\right\rangle \left\langle \alpha\right\vert
+b^{2}\left\vert \beta\right\rangle \left\langle \beta\right\vert
\otimes\left\vert \beta\right\rangle \left\langle \beta\right\vert
+abc\left\vert \alpha\right\rangle \left\langle \beta\right\vert
\otimes\left\vert \alpha\right\rangle \left\langle \beta\right\vert
+ab\overline{c}\left\vert \beta\right\rangle \left\langle \alpha\right\vert
\otimes\left\vert \beta\right\rangle \left\langle \alpha\right\vert ,
\end{align*}
where
\[
c=\sum_{i=1}^{k}c_{\alpha,i}\overline{c}_{\beta,i}.
\]
We claim that $c=0$. \ To see this, recall that $\Phi(\left\vert
\psi\right\rangle \left\langle \psi\right\vert )$\ is a separable
mixed state, and consider any decomposition of $\Phi(\left\vert
\psi\right\rangle \left\langle \psi\right\vert )$\ into separable
pure states. \ Since $\Phi(\left\vert \psi\right\rangle \left\langle
\psi\right\vert )$\ is a mixed state in the subspace spanned by
$\left\vert \alpha\right\rangle \left\vert \alpha\right\rangle $\
and $\left\vert \beta\right\rangle \left\vert \beta\right\rangle $,
every pure state\ in the support of $\Phi(\left\vert
\psi\right\rangle \left\langle \psi\right\vert )$\ must have the
form $x\left\vert \alpha\right\rangle \left\vert \alpha\right\rangle
+y\left\vert \beta\right\rangle \left\vert \beta\right\rangle $. \
But by the assumption $\left\vert \alpha\right\rangle \neq\left\vert
\beta\right\rangle $, such a state cannot be separable unless $x=0$\
or $y=0$. \ Hence the only separable pure states in the support of
$\Phi(\left\vert \psi\right\rangle \left\langle \psi\right\vert )$
are $\left\vert \alpha\right\rangle \left\vert \alpha\right\rangle $
and $\left\vert \beta\right\rangle \left\vert \beta\right\rangle $.
\ Therefore $abc=0$. \ But $a$ and $b$ were nonzero, so $c=0$ as
claimed.

This means in particular that $\Phi\left(  \left\vert
\phi_{\alpha}\right\rangle \left\langle \phi_{\beta}\right\vert
\right) =0$\ for all $\left\vert \alpha\right\rangle \neq\left\vert
\beta\right\rangle
$. \ Hence%
\begin{align*}
\langle\phi_{\beta}|\phi_{\alpha}\rangle &  =\sum_{i=1}^{k}\left\langle
\phi_{\beta}\right\vert E_{i}^{\dagger}E_{i}\left\vert \phi_{\alpha
}\right\rangle \\
&  =\operatorname*{Tr}\left(  \sum_{i=1}^{k}E_{i}\left\vert \phi_{\alpha
}\right\rangle \left\langle \phi_{\beta}\right\vert E_{i}^{\dagger}\right) \\
&  =\operatorname*{Tr}\left(  \Phi(\left\vert \phi_{\alpha}\right\rangle
\left\langle \phi_{\beta}\right\vert )\right) \\
&  =0.
\end{align*}
So for different $\left\vert \alpha\right\rangle $'s, the states $\left\vert
\phi_{\alpha}\right\rangle $ are all orthogonal, and since the number of
$\left\vert \alpha\right\rangle $'s is infinite, $\dim\mathcal{H}$ must be
infinite as well.
\end{proof}

\section{Open Problems\label{OPEN}}

\subsection{The Power of Multiple Merlins}

The power of $\mathsf{QMA}\left(  2\right)  $\ and related classes is still
poorly understood. \ Can we find a \textquotedblleft
classical\textquotedblright\ problem (for example, a group-theoretic problem
like those of Watrous \cite{watrous}) that is in $\mathsf{QMA}\left(
2\right)  $\ but not obviously in $\mathsf{QMA}$? \ Can we find a natural
$\mathsf{QMA}\left(  k\right)  $-complete\ promise problem?

We still do not have any upper bound on $\mathsf{QMA}\left(  2\right)
$\ better than the trivial $\mathsf{NEXP}$, or even good evidence for such an
upper bound. \ As mentioned before, an earlier version of this paper showed
that $\mathsf{QMA}\left(  2\right)  \subseteq\mathsf{PSPACE}$\ assuming the
\textquotedblleft Strong Amplification Conjecture,\textquotedblright\ but
Fernando Brandao (personal communication) subsequently showed that the same
conjecture implies $\mathsf{QMA}\left(  2\right)  =\mathsf{QMA}$. \ Can we
show under a plausible conjecture that $\mathsf{QMA}\left(  2\right)
\subseteq\mathsf{EXP}$ or (better yet) $\mathsf{QMA}\left(  2\right)
\subseteq\mathsf{PSPACE}$?

Regarding our \textsc{3Sat} protocol, can we reduce the number of provers to
two? \ Can we reduce the number of qubits below $\widetilde{O}\left(  \sqrt
{n}\right)  $, or\ alternatively, give some sort of evidence against this
possibility? \ For example, can we show that $\Omega\left(  \sqrt{n}\right)
$\ qubits are information-theoretically required for the Uniformity Test?
\ Also, can we show that a $\mathsf{QMA}\left(  k\right)  $\ protocol for
\textsc{3Sat}\ with $n^{o\left(  1\right)  }$\ qubits would have unlikely
complexity consequences?

A long-shot possibility would be to give a quantum algorithm to \textit{find}
the unentangled witnesses in the \textsc{3Sat}\ protocol, in as much time as
it would take were the witnesses entangled. \ This would yield a
$2^{\widetilde{O}\left(  \sqrt{n}\right)  }$-time quantum algorithm for
\textsc{3Sat}.

\subsection{Amplification and Other Complexity Issues}

In defining $\mathsf{QMA}\left(  k\right)  $, does it matter if the amplitudes
are reals or complex numbers? \ For $\mathsf{BQP}$\ and $\mathsf{QMA}$, it is
not hard to show that this distinction is irrelevant. \ Interestingly, though,
the usual equivalence proofs break down for $\mathsf{QMA}\left(  k\right)  $.
\ As evidence that $\mathsf{QMA}\left(  k\right)  $\ might actually be
sensitive to the difference between reals and complex numbers, consider the
analysis of our\ \textsc{3Sat} protocol: in Section \ref{MATCH}, the result
that we need becomes much simpler to prove when we assume all amplitudes are real.

Can we show directly (i.e., without proving the Weak Additivity
Conjecture) that $\mathsf{QMA}\left(  k\right)  =\mathsf{QMA}\left(
2\right)  $, or that $\mathsf{QMA}\left(  2\right)  $\ protocols can
be amplified?

Can we prove Conjecture \ref{gdmconj}: that there is no $\left(
\varepsilon,\delta\right)  $-disentangler with $\operatorname*{poly}\left(
n\right)  $\ qubits and $\varepsilon,\delta>0$? \ Can we at least rule out
such a disentangler when either $\varepsilon>0$\ \textit{or} $\delta>0$?
\ Related to that, can we give a quantum oracle\ $U$ (as defined by Aaronson
and Kuperberg \cite{ak}) such that $\mathsf{QMA}^{U}\neq\mathsf{QMA}%
^{U}\left(  2\right)  $? \ Can we at least show that Conjecture \ref{gdmconj}%
\ would imply the existence of such an oracle?

Define the complexity class $\mathsf{QMA}\left(  2;h\right)  $\ to be the same
as $\mathsf{QMA}\left(  2\right)  $, except that now, instead of being
completely unentangled, the two Merlins are allowed to share $h$ EPR pairs
$\frac{1}{\sqrt{2}}\left(  \left\vert 00\right\rangle +\left\vert
11\right\rangle \right)  $.\footnote{A similar notion was considered by
Gavinsky \cite{gavinsky}. \ By contrast, Kobayashi and Matsumoto
\cite{km:qmip} allow two provers to share an \textit{arbitrary} polynomial
amount of entanglement.} \ Can we show---perhaps under some assumption---that
$\mathsf{QMA}\left(  2;h\right)  =\mathsf{QMA}\left(  2\right)  $ for all
polynomial $h$?

\subsection{$\mathsf{QMA}\left(  k\right)  $ With Unentangled Measurements}

Recall that our \textsc{3Sat}\ protocol\ involved three tests: Satisfiability,
Symmetry, and Uniformity. \ Suppose we are willing to settle for completeness
$1-\varepsilon$\ rather than $1$, and suppose we modify the Uniformity Test so
that Arthur rejects on not seeing enough collisions. \ Then can the Symmetry
Test be omitted? \ If so, then the resulting protocol would have the extremely
interesting property of making no entangled measurements, yet nevertheless
depending crucially on the absence of entanglement among the witnesses.

More generally, define $\mathsf{BellQMA}\left(  k\right)  $\ to be
the subclass of $\mathsf{QMA}\left(  k\right)$\ in which Arthur is
restricted to making a separate measurement on each witness
$\left\vert \varphi _{i}\right\rangle $, with no entanglement
between the measurements, and no measurement depending on the
outcome of any other. \ (The name arises because Arthur is
essentially restricted to performing a \textquotedblleft Bell
experiment.\textquotedblright) \ In an earlier version of this
paper, we raised the question of whether $\mathsf{BellQMA}\left(
k\right)=\mathsf{QMA}$: that is, whether a $\mathsf{QMA}\left(
k\right)$ protocol with separate measurements on each witness ever
does superpolynomially better than an ordinary $\mathsf{QMA}$
protocol. \ Recently, Brandao \cite{brandao} managed to settle our
question, showing that $\mathsf{BellQMA}\left(
k\right)=\mathsf{QMA}$ for all constant $k$.\footnote{The case of
larger $k$ remains open, and therefore Brandao's result has no
implications for the earlier question about our \textsc{3Sat}\
protocol.} \ Indeed, Brandao's result applies even to the stronger
model in which Arthur can first measure $\left\vert \varphi
_{1}\right\rangle$, then choose a measurement for $\left\vert
\varphi _{2}\right\rangle$ depending on the outcome of the
$\left\vert \varphi _{1}\right\rangle$ measurement, and so on up to
$\left\vert \varphi _{k}\right\rangle$. \ (In other words, in which
there is one-way communication from earlier to later measurement
steps.)

On the other hand, define $\mathsf{LOCCQMA\left(  k\right)}$ to be
the subclass of $\mathsf{QMA}\left(  k\right)$\ in which Arthur's
measurements on the $k$ witnesses must be unentangled, but can
involve arbitrary local operations and classical communication. \
(In other words, Arthur can perform a partial measurement on
$\left\vert \varphi _{1}\right\rangle$, then based on the outcome of
that measurement perform a partial measurement on $\left\vert
\varphi _{2}\right\rangle$, then based on the outcome of that
measurement perform another partial measurement on $\left\vert
\varphi _{1}\right\rangle$, and so on.) \ Then where
$\mathsf{LOCCQMA\left(  k\right)}$ sits between $\mathsf{QMA}$ and
$\mathsf{QMA}\left(k\right)$ remains an open problem. \ (Note that
it is trivial to show amplification for $\mathsf{LOCCQMA}\left(
k\right) $, as for $\mathsf{BellQMA}\left( k\right) $. \ This is
because, without entangling measurements, the entanglement-swapping
problem described in Section \ref{AMPINT}\ can never arise.)

\section*{Acknowledgments}

We thank Fernando Brandao for pointing out that our
\textquotedblleft Strong Amplification Conjecture\textquotedblright\
implies not only
$\mathsf{QMA}\left(  2\right)  \subseteq\mathsf{PSPACE}$\ but $\mathsf{QMA}%
\left(  2\right)  =\mathsf{QMA}$, and that our original version of
Lemma \ref{2nlem}\ contained a bug, among other extremely helpful
comments; Norbert Schuch for pointing out that Theorem \ref{ampthm}\
can be based on a conjecture about entanglement of formation rather
than squashed entanglement; Madhu Sudan for pointers on the PCP
Theorem; John Watrous for posing Conjecture \ref{gdmconj}; an
anonymous reviewer for greatly simplifying the proof of Lemma
\ref{2nlem}\ and other help; and Patrick Hayden, Ryan O'Donnell,
Alain Tapp, and Andreas Winter for helpful discussions.

\bibliographystyle{plain}
\bibliography{thesis}

\end{document}